\documentclass[useAMS,usenatbib]{mn2e}
\usepackage{times}
\usepackage{txfonts}
\usepackage{bm}

\usepackage{color}
\usepackage[pdftex]{graphicx}
\usepackage[pdftex]{hyperref}

\newcommand\plotone[1]{\centering\includegraphics[width=\hsize]{#1}}

\newcommand\kms{{\rm \,km\,s^{-1}}}
\newcommand\jun{{\rm \,kpc}\,\kms}
\newcommand\beq{\begin{equation}}
\newcommand\eeq{\end{equation}}
\newcommand\vR{\rm v_R}
\newcommand\vphi{\rm v_\phi}
\newcommand\vr{\rm v_r}
\newcommand\vz{\rm v_z}
\newcommand\vth{\rm v_\theta}
\newcommand\rx{\rm x}
\newcommand\ry{\rm y}
\newcommand\rz{\rm z}
\newcommand\vx{\rm v_x}
\newcommand\vy{\rm v_y}
\newcommand\jz{\rm J_z}
\newcommand\jperp{\rm J_\perp}

\newcommand\masyr{${\rm mas\,yr}^{-1}$}
\newcommand\feh{[{\rm Fe/H}]}
\newcommand\D{{\rm D}}
\newcommand\sko{SKO}

\begin{document}

\title[Kinematics of SDSS subdwarfs]{Kinematics of SDSS subdwarfs:
  Structure and substructure of the Milky Way halo}

\author[Smith et al.]{
M.C. Smith$^1$\thanks{msmith@ast.cam.ac.uk}, 
N.W. Evans$^1$, V. Belokurov$^1$, P.C. Hewett$^1$, D.M. Bramich$^2$, 
    G. Gilmore$^1$, \newauthor
    M.J. Irwin$^1$, S. Vidrih$^{1,3,4}$, D.B. Zucker$^{1,5,6}$
 \medskip
 \\$^1$Institute of Astronomy, University of Cambridge, Madingley Road, Cambridge, CB3 0HA, UK
 \\$^2$Isaac Newton Group of Telescopes, Apartado de Correos 321,
 E-38700 Santa Cruz de la Palma, Canary Islands, Spain
 \\$^3$Astronomisches Rechen-Institut, Zentrum f\"ur Astronomie der 
 Universit\"at Heidelberg, M\"onchhofstrasse 12-14, 69120 Heidelberg,
 Germany
 \\$^4$Faculty of Mathematics and Physics, University of Ljubljana,
Ljubljana, Slovenia
 \\$^5$Department of Physics, Macquarie University, North Ryde, NSW 2109,
Australia
 \\$^6$Anglo-Australian Observatory, P. O. Box 296, Epping, NSW 1710, Australia}

\date{Accepted ........Received .......;in original form ......}

\pubyear{2008}

\maketitle

\begin{abstract}
  We construct a new sample of $\sim1700$ solar neighbourhood halo
  subdwarfs from the Sloan Digital Sky Survey, selected using a
  reduced proper motion diagram. Radial velocities come from the SDSS
  spectra and proper motions from the light-motion curve catalogue of
  \citet{Br08}. Using a photometric parallax relation to estimate
  distances gives us the full phase-space coordinates.  Typical
  velocity errors are in the range $30 - 50 \kms$.  This halo sample
  is one of the largest constructed to-date and the disc contamination
  is at a level of $\la$ 1 per cent.  This enables us to calculate
  the halo velocity dispersion to excellent accuracy. We find that the
  velocity dispersion tensor is aligned in spherical polar coordinates
  and that $(\sigma_{\rm r},\sigma_\phi,\sigma_\theta)$ = (143 $\pm$
  2, 82 $\pm$ 2, 77 $\pm$ 2) $\kms$. The stellar halo exhibits no net
  rotation, although the distribution of $\vphi$ shows tentative
  evidence for asymmetry.  The kinematics are consistent with a mildly
  flattened stellar density falling with distance like $r^{-3.75}$.

  Using the full phase-space coordinates, we look for signs of
  kinematic substructure in the stellar halo. We find evidence for
  four discrete overdensities localised in angular momentum and
  suggest that they may be possible accretion remnants. The most
  prominent is the solar neighbourhood stream previously identified by
  \citet{He99}, but the remaining three are new. One of these
  overdensities is potentially associated with a group of four
  globular clusters (NGC5466, NGC6934, M2 and M13) and raises the
  possibility that these could have been accreted as part of a much
  larger progenitor.
\end{abstract}


\section{Introduction}
\label{sec:intro}

In recent years, there has been an impressive and growing body of
evidence that the stellar halo of the Galaxy is composed of the
remnants of accretion and merging events.  The Sloan Digital Sky
Survey~\citep[SDSS;][]{Yo00} has made a significant contribution to
this field, with a series of major discoveries, including the
Monoceros Ring~\citep{Ne02,Ya03}, the Virgo Overdensity~\citep{Ju08}
and the Hercules-Aquila Cloud~\citep{Be07}. These structures were
identified as overdensities of resolved stars, exploiting the deep and
homogeneous photometry in 5 bands ($u,g,r,i$ and $z$) that SDSS
provides in a large area around the North Galactic Cap.

The aim of this paper is to search for substructure in the Galactic
halo with SDSS data, but this time using kinematic methods.  This
technique led to the discovery of the disrupting Sagittarius dwarf
galaxy and its Stream, which were found serendipitously in a radial
velocity survey of the outer Galactic Bulge~\citep{Ib94,Ib95}.  It has
also led to the identification of a number of probable halo streams
found by spectroscopic surveys of the blue horizontal branch
population~\citep[e.g.,][]{Ar92,Cl05}.  More rarely, searches for
kinematic substructure have exploited both proper motion and radial
velocities. For example, high quality proper motions provided by
HIPPARCOS, together with ground-based radial velocities, enabled
\citet[][ hereafter H99]{He99} to construct three-dimensional
velocity distributions for an almost complete sample of nearby halo
stars and to identify a nearby stream as a coherent structure in
velocity space.

Samples of accurate proper motions can be obtained through repeated
astrometric observations over a significant temporal baseline.
Most of the SDSS data has limited variability
information, with one important exception: during the three months
when the Southern Galactic Cap is available for observation, SDSS
repeatedly scanned a $\sim 290$ square degree area -- known as Stripe
82 -- to detect supernovae \citep[see e.g.,][]{Ab09}.

\citet{Br08} have presented a Stripe 82 catalogue of almost 4 million
``light-motion curves'', in which objects are matched between the
$\sim 30$ epochs, taking into account the effects of any proper motion
over the eight-year baseline.  The catalogue is complete down to a
magnitude 21.5 in $u, g, r$ and $i$, and to magnitude 20.5 in
$z$. Each object has its proper motion calculated based only on the
multi-epoch SDSS J2000 astrometric measurements. It reaches almost 2
magnitudes fainter than the SDSS/USNOB catalogue~\citep{Mo03,Mu04},
making it the deepest large-area photometric and astrometric catalogue
available.

Substructure in the catalogue of \citet{Br08} has already been
identified by looking for overdensities. For example, \citet{Wa09}
isolated the RR Lyrae variables by a combination of colour,
metallicity and period cuts. The Hercules-Aquila Cloud and the
Sagittarius Stream were both clearly identifiable in the Stripe 82 RR
Lyrae population, together with a completely new and distant
substructure called the Pisces Overdensity. Thus far, however,
searches for kinematic substructure in the Stripe 82 data have
not been carried out, and the object of this paper is to remedy this
deficiency.

In Section~\ref{sec:sample}, we show how to isolate a sample of nearby
halo subdwarfs from the catalogue of \citet{Br08}. The kinematic
properties of the local halo subdwarf population are discussed in
Section~\ref{sec:kinematics}. We develop algorithms to search for
kinematic substructure, recovering the known H99 stream in
Section~\ref{sec:substr}, as well as new kinematic overdensities in
Section~\ref{sec:new}.

\begin{table}
\begin{tabular}{llc}
\hline Section & Description of cut & Number of objects\\
\hline
\ref{sec:rpm} & Initial sample with high quality astrometry & 370,000\\ 
\ref{sec:rpm} & Satisfy RPM cut & 27,000\\ 
\ref{sec:spectra} & With high quality spectra & 2,704\\ 
\ref{sec:dist} & Colour within range of photometric parallax & 2,183\\ 
\ref{sec:dist} & Distance $<$ 5.0 kpc & 1,717\\ 
\hline
\end{tabular}
\caption{The number of subdwarf candidates after the respective cuts.}
\label{tab:sample_num}
\end{table}

\section{The subdwarf sample}
\label{sec:sample}

\subsection{Reduced proper motion (RPM) diagram}
\label{sec:rpm}

\begin{figure}
\plotone{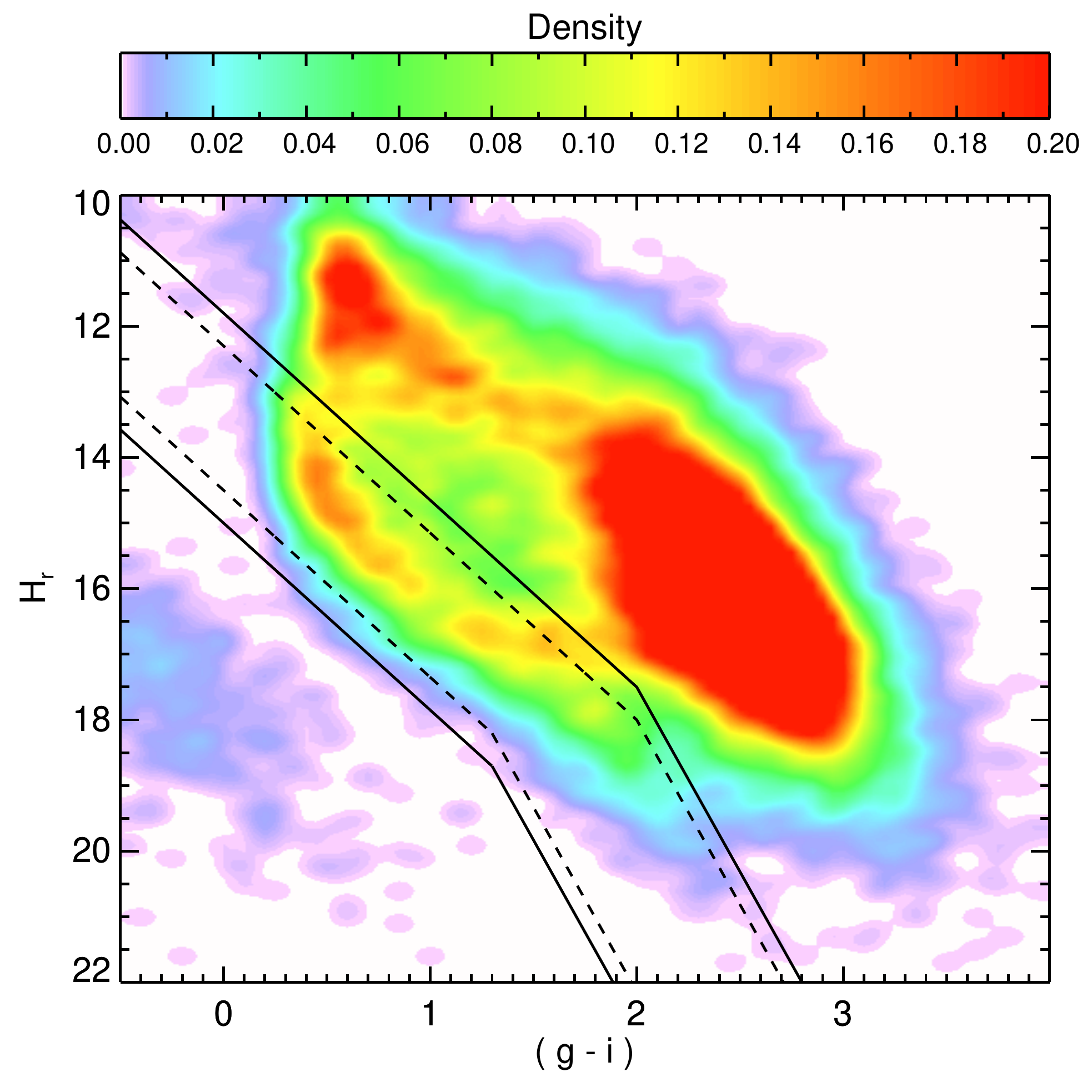}
\caption{Reduced proper motion diagram for the \citet{Br08}
  sample. The solid lines denote the approximate boundary of the
  subdwarf population. The dashed lines denote the stricter boundary
  employed to reduce contamination from disc main-sequence dwarfs
  (which lie to the upper-right of the boundary) or white dwarfs
  (which lie to the lower-left). To aid clarity in this figure, we
  have imposed an additional proper motion cut ($\mu > 30$ \masyr); however, no such cut
  was applied to our sample. Note that the colour-scale saturates at
  20 per cent of the peak density.}
\label{fig:rpm}
\end{figure}

The full \citet{Br08} catalogue contains proper motions, $\mu$, for $\sim 1$ million
stars down to magnitudes of $r \sim 21.5$. We trim this sample by
enforcing the following three conditions to retain only those objects
with high quality data. First, we insist that the mean
object type $\ge$ 5.7. This requires an object to be classified as a
star in $\ge$ 90 per cent of epochs. This is less than 100 per cent in
order to retain objects that have been misclassified in a limited
number of epochs due to problems with the SDSS star-galaxy separation
algorithm. In the \citet{Br08} catalogue, this occurs particularly in
the final season, when observations were not exclusively taken in
photometric conditions. Secondly, we insist that the proper-motion
error is less than 4 \masyr to remove stars with poorly determined
proper motions.
All of our final sample have proper motions based on at least 23
epochs and time-baselines of over 4 years, which indicates that the
formal proper motion errors should be reliable.
Thirdly, we impose $r < 19.5$ to reduce contaminants (see Appendix A
for more details). Here, and elsewhere in the paper, all magnitudes
are corrected for extinction using the maps of \citet{Sc98}. The
magnitude cut is often redundant as we also later require stars to
have SDSS spectra, and there are very few spectra for stars fainter
than this limit.

We select our candidate subdwarfs from a reduced proper motion (RPM)
diagram. The $r$-band RPM is given by
\beq
H_r = r + 5 {\rm log} \frac{\mu}{{\rm mas\,yr}^{-1}} - 10,
\eeq
where $\mu$ is the proper-motion and $r$ is the apparent magnitude.
This uses a star's proper motion as a proxy for distance, allowing us
to separate cleanly populations with different absolute magnitudes
(e.g. main sequence dwarfs, white dwarfs, giants).
Although disc and halo dwarfs have similar absolute
magnitudes, they have very different kinematics. As a consequence, the
faster moving halo stars appear offset from the dominant disc stars in
the RPM diagram, which can be seen clearly in our data (Fig~\ref{fig:rpm}).

Note that for the purposes of this figure, we have enforced an
additional cut that $\mu > 30$ \masyr{} to emphasise the distinction
between disc and halo dwarfs. As we relax this criterion, the diagram
becomes populated by slower moving halo dwarfs which lie in the
boundary region between the two populations on this RPM
diagram. Slower moving disc dwarfs, on the other hand, lie in the
upper portion of the diagram well away from our cuts and hence do
not contaminate our final sample. Therefore, although placing such a
cut on the magnitude of $\mu$ aids the clarity of this figure, it is
not necessary for our final sample. By not placing a cut on $\mu$, we
make the process of quantifying the kinematic bias much easier (as
described later in Section \ref{sec:bias}).

We now define our halo subdwarf region as
\begin{eqnarray}
 H < 2.85 (g\!-\!i) + 11.8 & {\rm for} & (g\!-\!i) \le 2 \nonumber \\
 H < 5.63 (g\!-\!i) + 6.24 & {\rm for} & (g\!-\!i) > 2 \nonumber \\
 H > 2.85 (g\!-\!i) + 15.0 & {\rm for} & (g\!-\!i) \le 1.3 \nonumber \\
 H > 5.63 (g\!-\!i) + 11.386 & {\rm for} & (g\!-\!i) > 1.3.
\end{eqnarray}
To construct as clean a sample as possible, we reject all objects that
lie within $\Delta H = 0.5$ of the boundary, resulting in the
final region shown by the dashed line in Fig. \ref{fig:rpm}.

\subsection{SDSS spectra}
\label{sec:spectra}

As well as its photometric survey, the 7th SDSS Data Release
\citep[DR7; ][]{Ab09} also has a large number of spectra. These are
predominantly of galaxies and quasars, but also include stellar
targets \citep{Ya09}. These have been analysed by a pipeline designed
to derive the radial velocities and fundamental stellar atmospheric
parameters \citep[the SEGUE Spectral Parameter Pipeline, or SSPP;
][]{Le08a, Le08b, Pr08}.  Typical internal errors reported by the
pipeline are $\delta {\rm v_r} \sim 3.8 \kms$ for the radial velocity
and $\delta \feh\ \sim 0.1$ dex for the metallicity. Validation of
this pipeline was carried out through comparison with high resolution
spectra \citep{Pr08}, which showed that the SSPP external errors are
approximately 2.4 $\kms$ in velocity and 0.11 dex in metallicity. To
account for this, we add these errors in quadrature to the internal
errors from the SSPP.

We have cross-matched our candidate subdwarf sample with the SSPP DR7
catalogue, and found that $\sim 7$ per cent have suitable spectra --
defined as those with SSPP flag set to `nnnn', which indicates that
there are no cautionary signs and the stellar parameters should be
well determined \citep{Le08a}. Additionally, we reject all spectra
with radial velocity error greater than $50 \kms$. If there are
multiple spectra, we take the spectrum with greatest signal-to-noise
ratio. The median errors in our cross-matched sample are $\delta {\rm
  v_r} = 4.1\kms$ and $\delta {\rm [Fe/H]} = 0.12$ dex.

The SDSS spectroscopic target selection function is very
heterogeneous, covering a wide variety of targets
\citep{Ab09,Ya09}. This will introduce selection effects into
our sample, for example much of the stellar targeting is biased
towards bluer populations which are likely to have lower
metallicities. However, we do not believe that 
this will introduce any significant kinematic bias. If there was a
strong correlation between the kinematics and metallicity $within$ the
halo subdwarf population, then our results would be kinematically
biased. However, as we will see later in Section \ref{sec:kinematics},
we do not find any such strong correlation in our sample.

\begin{figure}
\plotone{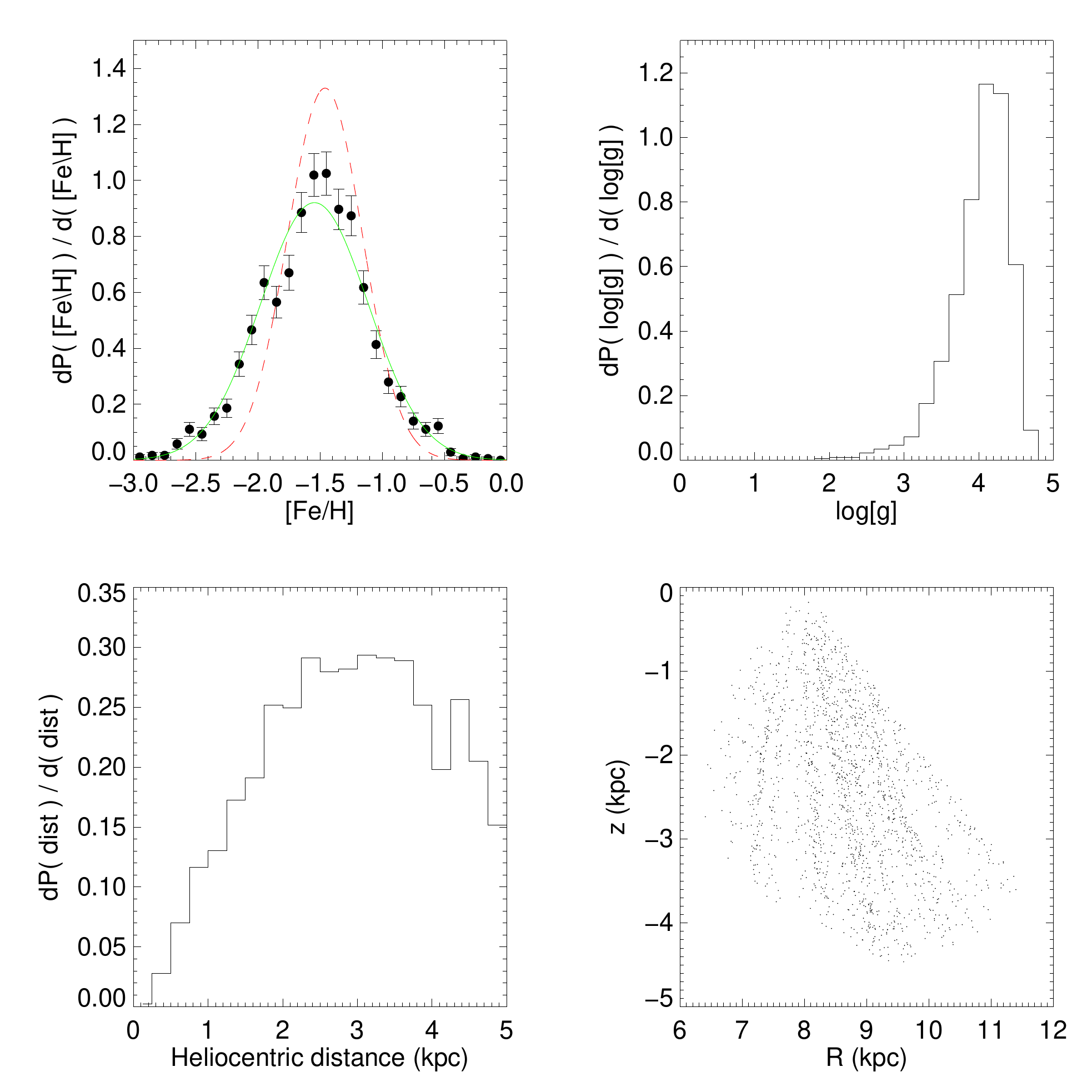}
\caption{Various properties of our subdwarf sample, namely metallicity
  (top left), surface gravity (top right), heliocentric distance
  (bottom left) and spatial distribution in Galactocentric cylindrical
  coordinates (bottom right). Our $\feh$ distribution may be subject
  to biases due to the nature of the sample construction (see Section
  \ref{sec:dist}). For comparison, the dashed curve shows the
  distribution from \citet{Iv08}. The spatial distribution is
  inhomogeneous due to the fact that the proper motion accuracy, which
  depends on the number of epochs that a star was observed, varies
  significantly across our field of view \citep[see][]{Br08}.}
\label{fig:non_kin}
\end{figure}

\subsection{Distances, metallicities and surface gravities}
\label{sec:dist}

To obtain distances for our subdwarfs, we use a photometric parallax
relation. Such a relation has been established for main-sequence stars
in the SDSS photometric system by \citet{Iv08}. It was constructed
through observations of eleven globular clusters and tested using
additional clusters and stars with trigonometric parallax distances.

The relation, which is given in equations (A1-A5) of \citet{Iv08}, is
valid for dwarfs with $0.2 < g\!-\!i < 4$ and incorporates a
correction to account for systematic trends with
metallicity. According to \citet{Iv08}, the intrinsic scatter in this
relation is $\sim 0.13$ mag, which is a lower limit on the true
uncertainty. An upper limit of $0.3$ has been estimated by
\citet[][see also Juri\'c et al. 2008]{Se08}, who use candidate binary
stars to assess the reliability of the photometric
parallaxes. Therefore, for this work, we take a fiducial uncertainty
of 0.2 mag.

The relation we adopt differs from that of \citet{Iv08} in that we
have chosen not to incorporate their turn-off correction (given by
equation A6 of their paper). Instead, we have
constructed our own correction using the stellar models of
\citet{Do08}. Our procedure, which is described in Appendix
\ref{app:turn-off}, results in the following correction for stars
with $0.3 < g\!-\!i < 0.6$,
\beq
\label{eq:turn-off}
\Delta M_r = 
a_0\,x + 
a_1\,x\,y +
a_2\,x^3 +
a_3\,x^2\,y +
a_4\,x^3\,y,
\eeq
where $x = (g\!-\!i) - 0.6$, $y = \feh$, and $a_0 = 2.87,\: a_1 =
2.25,\: a_2 = -9.79,\: a_3 = 2.07,\: a_4 = 0.31$.
Due to the scatter in this relation we increase the uncertainties in
the distances of the stars in this colour range (see Appendix
\ref{app:turn-off}).

Using the accurate photometry from the multi-epoch Stripe 82 data
(with mean $\delta(g - i) < 0.01$) and taking \feh\ from the spectra,
we now estimate a distance. Combining the uncertainty in the
\citet{Iv08} relation with the error on \feh, we find that the median
distance error in our sample is $\sim 11$ per cent. Our distances are
mostly in the range 0.5 to 5 kpc, although there are a small number of stars
further out. However, unless otherwise stated, henceforth we only use
stars with heliocentric distance $\D < 5$ kpc. Beyond this, the errors
on the velocities become too large because of the uncertainty in the
proper motion. This gives us a final sample of 1717 halo subdwarfs, as
summarized in Table~\ref{tab:sample_num}.

There are a number of potential sources of contamination in the
sample, including white dwarfs, thin and thick disc stars and
background giants. These are discussed in turn in Appendix A, which
concludes that the actual level of contamination is very small ($\la$
1 per cent).  As a check that our sample is clean, we examine
various non-kinematic properties in Fig. \ref{fig:non_kin}.  We can
immediately see that the surface gravities are consistent with that of
a dwarf population. Although there are a handful of stars with low
gravity, the vast majority have log(g) $>$ 3. These outliers are most
likely caused by misestimation of log(g) from the spectra. The median
error on log(g) for our sample is 0.25 dex, but the median error for
those stars with log(g)$<$3 is 0.5 dex.

The distribution of \feh\ is also consistent with a halo sample. In
Fig. \ref{fig:non_kin}, we show the observed distribution along with
the Gaussian found by \citet{Iv08} using photometric
metallicities. Although the mean of our distribution ($\feh = -1.55$)
is similar to theirs ($\feh = -1.46$), it is clear that our
distribution is significantly broader (with a dispersion 0.43 as
opposed to 0.3). This broadening and offset is most likely due to the
underlying selection function for the SEGUE spectra, which will
clearly have a significant effect on the \feh\ distribution.

\begin{figure*}
\centering\includegraphics[width=\hsize]
{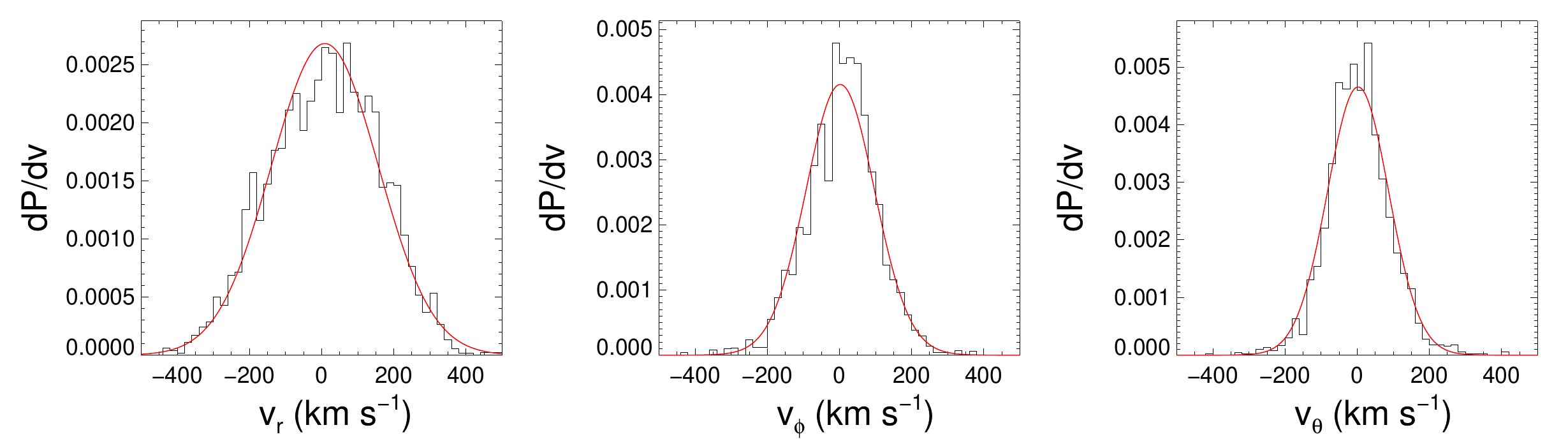}
\caption{The velocity distributions of the halo subdwarf sample. The
  velocity components are resolved in spherical polar coordinates with prograde
  rotation corresponding to $\vphi < 0$. The solid curves show
  Gaussian fits to the data, with the corresponding parameters given
  in Table \ref{tab:uvw}.}
\label{fig:uvw}
\end{figure*}
\begin{figure}
\plotone{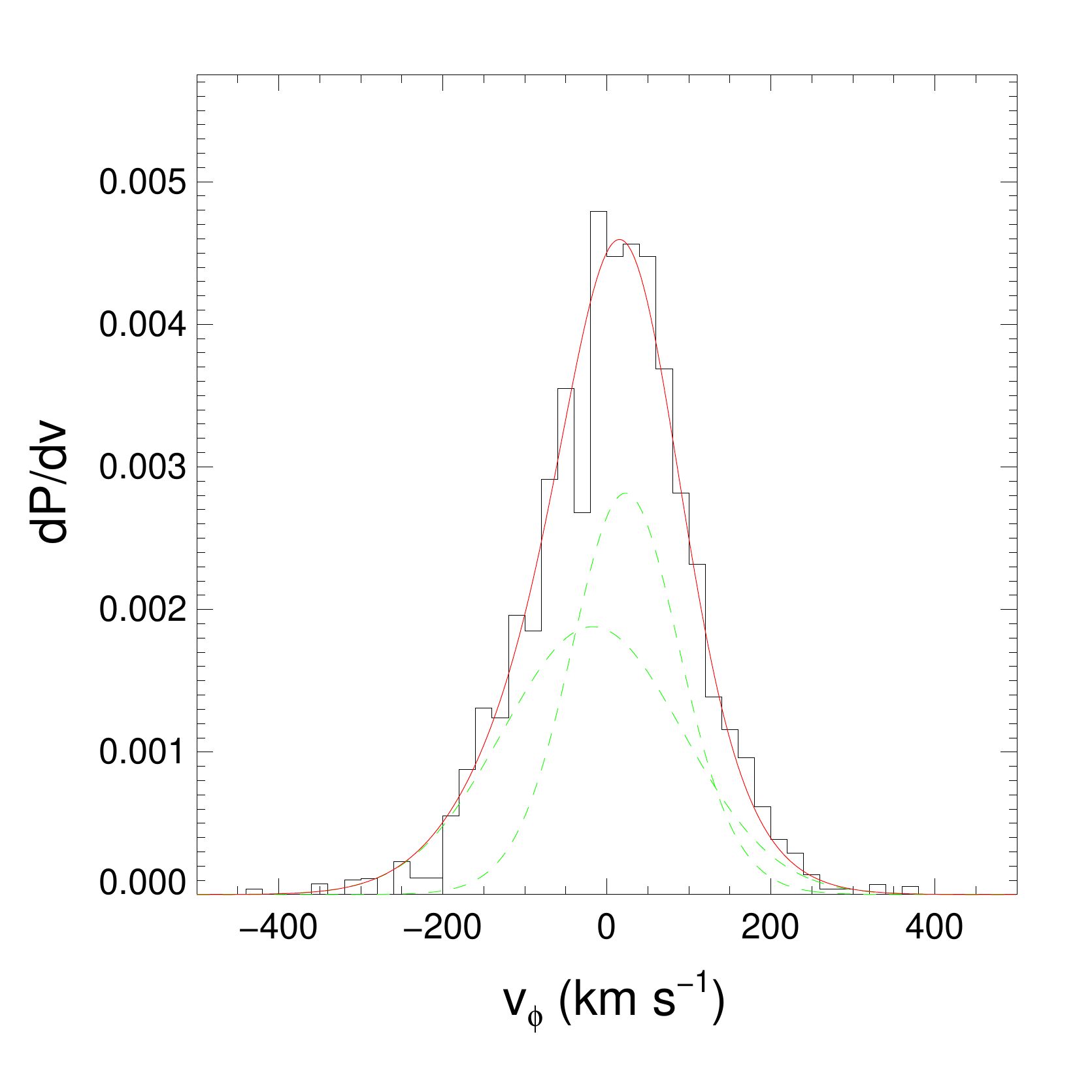}
\caption{Multi-component Gaussian fit to the $\vphi$ velocity
  distribution. The statistically preferred fit is this two-component
  model, which is better able to reproduce the asymmetry than the
  single Gaussian shown in Fig. \ref{fig:uvw}.}
\label{fig:vphi}
\end{figure}

\begin{table*}
\begin{tabular}{ccccc}
\hline
System & Component & $\left< {\rm v} \right> \;\; (\kms)$ 
& $\sigma_{\rm v}\;\;(\kms)$ \\
\hline
Spherical &$\vr$ & $8.9^{+2.8}_{-2.6}$ & $142.7^{+2.0}_{-1.7}$ \\
&$\vphi$ & $2.3^{+1.6}_{-1.8}$ & $82.4 \pm 1.4$\\
&$\vth$ & $2.4^{+1.6}_{-1.4}$ & $77.3 \pm 1.1$\\
\\
Cylindrical &$\vR$ & $8.9 \pm 2.6$ & $138.2^{+2.0}_{-1.7}$\\
&$\vz$ & $-1.2 \pm 1.6$ & $89.3^{+1.3}_{-1.1}$\\
\hline
\end{tabular}
\caption{Efficiency weighted mean velocities and their dispersions for
  our halo sample in both spherical and cylindrical polar
  coordinates. The values are obtained using a standard maximum 
  likelihood technique which incorporates the spread that results 
  from observational errors (see equation~\ref{eq:lhood}). The 
  cross-terms or covariances are reported in a separate 
  paper~\citep{Sm09}.}
\label{tab:uvw}
\end{table*}

\begin{table*}
\begin{tabular}{ccccc}
\hline
Reference & System & Component & $\left< {\rm v} \right> \;\; (\kms)$ & $\sigma_{\rm v}\;\;(\kms)$\\
\hline
\citet{Wo78} & Spherical & $\vr$ & -- & 148 \\
&&$\vphi$ & -- & 122\\
&&$\vth$ & -- & 82\\\\
\citet{No86} & Cartesian & $\vx$ & -- & 131 $\pm$ 6 \\
&&$\vy$ & -- & 106 $\pm$ 6\\
&&$\vz$ & -- & 85 $\pm$ 4\\\\
\citet{Go98} & Cartesian & $\vx$ & -5 $\pm$ 15 & 171 $\pm$ 11 \\
(Kinematically selected sample) &&$\vy$ & 9 $\pm$ 13 & 99 $\pm$ 8\\
&&$\vz$ & 1 $\pm$ 8 & 90 $\pm$ 7\\\\
\citet{Go98} & Cartesian & $\vx$ & -3 $\pm$ 9 & 160 $\pm$ 7 \\
(Non-kinematically selected sample) &&$\vy$ & 28 $\pm$ 9 & 109 $\pm$ 9\\
&&$\vz$ & 2 $\pm$ 5 & 94 $\pm$ 5\\\\
\citet{Ch00} & Cartesian & $\vx$ & -16 $\pm$ 16 & 141 $\pm$ 11 \\
&&$\vy$ & 26 $\pm$ 12 & 106 $\pm$ 9\\
&&$\vz$ & -5 $\pm$ 11 & 94 $\pm$ 8\\\\
\citet{Go03} & Cartesian & $\vx$ & -1 $\pm$ 3 & 171 $\pm$ 10 \\
&&$\vy$ & 9 (fixed) & 99 $\pm$ 8\\
&&$\vz$ & 2 $\pm$ 3 & 90 $\pm$ 7\\\\
\citet{Ke07} & Cylindrical & $\vR$ & -4 $\pm$ 11 & 157 $\pm$ 8 \\
&&$\vphi$ & -23 $\pm$ 8 & 110 $\pm$ 6\\
&&$\vz$ & -1 $\pm$ 6 & 84 $\pm$ 4\\\\
\hline
\end{tabular}
  \caption{Some determinations of the halo velocity parameters from
    the literature. Note that $\vphi$ is defined counter to the 
    direction of disc rotation and for nearby samples $\vphi 
    \approx -\vy$. With the exception of \citet{Ke07}, the estimates 
    for the bulk motion have been standardised using the value of the 
    solar motion from \citet{De98}. \citet{Ke07} use the value from 
    \citet{Mi81} and as a consequence their estimate of $\langle 
    \vphi \rangle $ is subject to an offset of $\sim7\kms$.}
\label{tab:uvw_lit}
\end{table*}

\begin{table*}
\begin{tabular}{ccc}
\hline
Component & ${\rm Corr}[{\rm v}_i,{\rm v}_j]$ & $\alpha_{i,j}\;(^\circ)$\\
\hline
$[{\rm v_r,v_\theta}]$& 0.052 $\pm$ 0.028 & 2.2 $\pm$ 1.2\\
$[{\rm v_r,v_\phi}]$& -0.019 $\pm$ 0.042 & -1.6 $\pm$ 3.6\\
$[{\rm v_\phi,v_\theta}]$& -0.073 $\pm$ 0.050 & -36.7 $\pm$ 24.1\\
\hline
\end{tabular}
\caption{Efficiency weighted correlation coefficients and tilt angles
  for halo stars with $\left|\,z\,\right| > 1$ kpc 
  \citep[see][for the definitions of these quantities]{Sm09}. Note that these
  values are not identical to those quoted in \citet{Sm09} because the ones
  given here incorporate our new turn-off correction for the
  photometric parallax relation (see equation
  \ref{eq:turn-off}). However, the difference is small and none of the
  conclusions of \citet{Sm09} are affected.}
\label{tab:tilt}
\end{table*}

\subsection{Kinematic bias}
\label{sec:bias}

One problem with our sample is that it suffers from kinematical
bias. This is caused by the cut on $H_{\rm r}$, which selects stars via
their tangential velocity, rather than their proper motion. Using a
cut on $H_{\rm r}$, instead of the proper motion, makes the task of
quantifying the bias significantly easier, since assumptions about the
underlying distance distribution (i.e., luminosity function) are not needed.

We calculate our detection efficiency as follows. For each subdwarf in
our final sample, we take the sky coordinates and create a mock sample
of 50,000 fake stars. Then, for each mock star, we select $M_r$ and
$g\! -\! i$ at random from our observed distribution. Note that for each
realization $both$ the magnitude and color are assigned from one star.
We then fix the kinematics by choosing velocities from Gaussians with
the halo velocity dispersions from \citet{Ke07} and no net motion
\citep[following][]{Al06}.  We determine the tangential component of
the velocity ($v_{\rm tan}$) with respect to the line-of-sight between
the Sun and the mock star. \footnote{To calculate ${\rm v_{tan}}$
  requires us to assign a distance to each mock star, which we do at
  random from the observed distribution. This means that the
  efficiency does have a dependence on the distance distribution;
  however this dependence is negligible since equation (\ref{eq:h_mock}) is
  a function of the tangential velocity rather than the proper
  motion.} Given this information we can calculate the reduced proper
motion using,
\beq
\label{eq:h_mock}
H_r = M_r + 5~{\rm log_{10}}\left( \frac{v_{\rm tan}}{4.74\kms}
\right).
\eeq
The efficiency is then given by the fraction of mock stars which pass
our reduced proper motion cut. 

\section{Global kinematic properties}
\label{sec:kinematics}

\subsection{Data}

Here, we calculate properties of the first and second moments of the
halo velocity distribution. We begin by defining our coordinate
systems.  We take the solar radius as 8 kpc, the velocity of the local
standard of rest as $220 \kms$ and the solar peculiar velocity as
given by \citet{De98}. The Galactocentric Cartesian reference frame is
denoted by ($\rx$, $\ry$, $\rz$), where the axes are oriented along
the line connecting the Sun and the Galactic centre, in the direction
of disc rotation and toward the North Galactic Pole,
respectively. This is a right handed frame so that the Sun is at $\rx
= - 8.0$ kpc. Velocity components resolved with respect to this
coordinate system are ($\vx$, $\vy$, $\vz$).

We also use cylindrical and spherical polar coordinates defined with
respect to a right-handed Galactocentric frame.  Cylindrical polars
are denoted by ($R, \phi, z$) where $R$ is radially outward, $\phi$ is
positive in the direction of counter-rotation of the disc and $z$ is
positive towards the North Galactic Pole. The corresponding velocity
components are ($\vR$, $\vphi$, $\vz$). Similarly, spherical polars
are denoted ($r, \theta,\phi$) where $r$ is radially outward and
$\theta$ is increasing towards the South Galactic Pole. The velocity
components are ($\vr$, $\vphi$, $\vth$). So, for stars in the solar
neighbourhood, $\vth \approx -\vz$. Note that disc stars rotate with
$\vphi \approx -220\kms$, which is a consequence of adopting
a right-handed system.

The values for the mean and dispersion of the velocity components in
spherical and cylindrical polars are given in Table~\ref{tab:uvw}.
They are calculated using a maximum likelihood technique which
corrects for the spread that results from observational errors,
namely,
\beq
\label{eq:lhood}
{\rm L}(\mu_i,\sigma_{v_i}) = \prod_{k=1}^N
\left\{
\frac{1}{\sqrt{2\pi\left(\sigma_i^2+\delta{\rm v}_{i,k}^2\right)}}
{\rm exp}
\left[
  \frac{-({\rm v}_{i,k} - \mu_i)^2}{2\left(\sigma_i^2+\delta{\rm v}_{i,k}^2\right)}
\right]
\right\}^{(1/\epsilon_k)},
\eeq
where $\mu_i$ and $\sigma_i$ are the mean and dispersion of velocity
component $i$, whilst ${\rm v}_{i,k}$ and $\delta{\rm v}_{i,k}$ are
the velocity and its uncertainty for the $k$th star, $\epsilon_k$ is
the detection efficiency for this star, and $N$ is the total number of
stars in the sample.  The errors on the individual velocities are
calculated using Monte Carlo methods, incorporating the errors on the
proper motion, distance and radial velocity. The median error in
$(\vr,\vphi,\vth)$ is (38, 47, 35) $\kms$. The magnitude of the errors
differs between the three components because of the location of our
field, which causes the proper motion to contribute more to the
$\vphi$ component. Furthermore, because the uncertainties are larger
in the tangential direction the errors are correlated, with the
orientation of the error ellipse varying along the field.

The maximum likelihood velocity distributions are displayed in
Fig.~\ref{fig:uvw}. Although the Gaussian models provide a reasonable
match to the observations for $\vr$ and $\vth$, it appears that
$\vphi$ exhibits some asymmetry.  Accordingly, we repeated the maximum
likelihood fitting allowing for multiple Gaussian components.  We
determine whether a multiple Gaussian model is statistically preferred
by comparing the ratio of likelihoods, where $-2 ~ {\rm ln}\left({\rm
    L}_i/{\rm L}_{i+1}\right) > 3$ indicates that a model with $(i+1)$
components is preferred. The value of 3 comes from the fact that there
are three additional parameters in the model with $(i+1)$ components.

Although for $\vr$ there is only a marginal improvement with multiple
Gaussians, for $\vth$ a two-component fit is clearly preferred,
reflecting the fact that the distribution is narrower than a single
Gaussian yet has noticeable wings for $|\vth| \ga 200 \kms$. As a
consequence, this two-component model has one main narrow component
(with $\sigma = 64\kms$) and a shallow broad component (with $\sigma =
135\kms$), both of which have mean close to 0 $\kms$. The presence of
these wings can be understood when one considers the substructures
present in our sample (which will be discussed later).

For $\vphi$, a two-component fit is also preferred, with approximately
equal contributions from the two components with dispersions of 48.1
and 98.9 $\kms$ and means 22.8 and -17.1 $\kms$, respectively (see
Fig. \ref{fig:vphi}). This provides a better match to the asymmetry in
the $\vphi$ distribution.  We believe that this asymmetry is not
caused by problems with our efficiency correction nor with
contamination from the disc (see Appendix A). This leads us to
conclude that it is a real effect. It appears that the asymmetry is
more pronounced for metal-rich stars ($\feh>-1.5$ dex), although we
find no evidence for dramatic trends with metallicity, such as those
postulated by \citet{Mo09}. Neither do we see any clear gradient in
$\langle\vphi\rangle$ as a function of height from the Galactic
plane. Although we identify various kinematic substructures in our
sample (see Sections \ref{sec:substr} and \ref{sec:new}), none of
these can explain this asymmetry.


In Table \ref{tab:uvw_lit}, we present earlier determinations of the
mean velocity and dispersion for halo stars in the literature.  Our
dispersions are significantly smaller than previous estimates
\citep[such as][]{Go98,Ch00,Go03,Ke07}, although the ratios are in
approximate agreement.  Comparison between different investigations is
clouded by the fact that some of the earlier samples are subject to
significant levels of disc contamination. Also, it is not always clear
whether corrections have been made for the spread induced by
measurement errors. The dispersions estimated by \citet{Go03} are
upper limits as they do not incorporate the spread due to
uncertainties in their photometric parallax relation.

There has also been controversy in the literature with regards to the
mean rotational velocity of the halo, with estimates of prograde
\citep{Ch00,Ke07}, no rotation \citep{Go98, Al06}, or retrograde
\citep{Ma92}. Our value of $\langle\vphi\rangle = 2.3^{+1.6}_{-1.8}
\kms$ is approximately consistent with a non-rotating halo. Since our kinematic
selection is biased against stars with $\vphi \approx -200 \kms$, we
might expect that our estimate of $\langle\vphi\rangle$ is dependent
on the efficiency correction. However, tests show that this is not the
case. Our calculation in Section \ref{sec:bias} requires an assumption
for the rotational velocity of the halo, but if we adopt values
anywhere between -20 and 20 $\kms$, then our measured value of
$\langle \vphi \rangle$ varies by less than 1 $\kms$. However, it is
still true that our estimate of $\langle\vphi\rangle$ (as with any of
the estimates from Table \ref{tab:uvw_lit}) is degenerate with the
assumed value for the local standard of rest ($220\kms$).

Another result of interest is the value of $\langle\vr\rangle$, which
is 3$\sigma$ away from zero. It is not clear why we obtain a positive
value of $\langle\vr\rangle$ (i.e., radially outward). There are a
variety of potential issues that could cause this effect: kinematic
substructure; systematic errors in the velocities (either the radial
velocities or proper motions) or distances; the presence of binary
stars, for which we would systematically underestimate their
distances; bias in the determination of the solar motion with respect
to the local standard of rest. We do not investigate this issue
further in the present paper.

We also calculate the efficiency weighted correlation coefficients and
tilt angles, as defined in \citet{Sm09}, for the 1,568 stars with
$\left|\,z\,\right| > 1$ kpc. Although the data used here
is the same as that of \citet{Sm09}, our sample differs in that we are
now incorporating our new turn-off correction to the parallax relation
(see equation \ref{eq:turn-off} and Appendix \ref{app:turn-off}). The
values are given in Table \ref{tab:tilt}. These are very similar to
those quoted in \citet{Sm09} and their conclusions are not affected
by this new parallax correction term.

\subsection{Modelling}

Taking our cue from the recent arguments of \citet{Fe06},
let us assume that the gravitational potential of the dark halo is
spherically symmetric. If the halo has a flat rotation curve of
amplitude $v_0$ (= 220 kms$^{-1}$), then the potential is
\begin{equation}
\psi = -v_0^2 \log r.
\end{equation}
We now seek a phase space distribution function for the stellar halo
that can reproduce the observed kinematics of the SDSS subdwarfs. From
Jeans Theorem \citep[see, e.g.,][]{Bi87}, the distribution
function must depend on the integrals of motion, namely the binding
energy $E$, the components of angular momentum, $J_x$, $J_y$, $J_z$,
together with the total angular momentum, $J$.

A number of authors \citep{Wh85,ZES} have shown that the stellar
density laws of the form
\begin{equation}
\rho = \rho_0 r^{-\gamma} \sin^{2n} \theta
\end{equation}
can be reproduced by distribution functions
\begin{equation}
  f_{m,n} (E,J^2,J_z^2) = \eta_{m, n} J^{2m} J_z^{2n} \exp [ ( \gamma \!+\! 2n\! +\! 2m) E ],
\end{equation}
where $m+ n > -1$ and $2n > -1$, and $\eta_{m, n}$ is a normalization
constant \citep[given in equation 3.5 of][]{ZES}. Note that we
have written $m$ and $n$ so that our notation is consistent with
earlier work, but $m$ and $n$ are not necessarily integers.

The corresponding velocity second moments are
\begin{eqnarray}
\rho \langle v_r^2 \rangle &=& {\rho_0 v_0^2 \over 2m + 2n + \gamma }
                               {\sin^{2n}\theta \over r^\gamma}, \nonumber\\
\rho \langle v_\theta^2 \rangle &=& { m + n + 1 \over n+1} \rho\langle v_r^2 
\rangle, \\
\rho\langle v_\phi^2 \rangle &=& (2n+1) \rho\langle
v_\theta^2 \rangle. \nonumber
\end{eqnarray}
So, for example, if $m=n=0$, the model is isotropic and all three
velocity dispersions are just $v_0/\sqrt{\gamma}$. This is a
well-known result for isothermal populations. More generally, for
fixed $\gamma$ and $n$, the velocity dispersions and the density have
the same angular dependence for all $m$, but the anisotropy ratios
$\langle v_\theta^2 \rangle/ \langle v_r^2 \rangle$ and $\langle
v_\phi^2 \rangle/\langle v_r^2 \rangle$ do depend on $m$.

A flexible way to model the stellar halo is to build further
distribution functions by linear superposition. Here, our aim is to
construct a simple distribution function that reproduces the
kinematics of the SDSS subdwarfs, and so we choose
\begin{equation}
f(E, J^2 , J_z^2) = \alpha_{m,0}f_{m,0} + \alpha_{m,1}f_{m,1},
\label{eq:twocomp}
\end{equation}
where the $\alpha_{m,n}$ are constants specifying the fraction
contributed by each component. The corresponding density law is
flattened and has the form
\begin{equation}
\rho(r,\theta) = \rho_0 r^{-\gamma} + \rho_1 r^{-\gamma} \sin^2 \theta.
\end{equation}
We require that at $R = 8.70$ kpc and $z = -2.41$ kpc (the centroid of
our SDSS subdwarf population)
\begin{equation}
\langle v_r^2 \rangle^{1/2} = 143 {\rm kms}^{-1},\quad
\langle v_\theta^2 \rangle^{1/2} = 77 {\rm kms}^{-1},\quad
\langle v_\phi^2 \rangle^{1/2} = 82 {\rm kms}^{-1}.
\end{equation}
The advantage of a two component model (\ref{eq:twocomp}) is that
there is a unique solution.  There are three velocity dispersion
constraints, and there are three unknowns, namely the radial fall-off
$\gamma$, the index $m$ and the ratio $\rho_1/\rho_0$ (or equivalently
$\alpha_{m,1}/\alpha_{m,0}$).  Solving the non-linear simultaneous
equations numerically gives the solution $\gamma = 3.75$, $m = -0.72$
and $\rho_1/\rho_0 = 0.063$.

In other words, the SDSS subdwarf kinematics are consistent with a
stellar halo in which the density falls off like $r^{-3.75}$, somewhat
steeper than the $r^{-3.5}$ advocated in the classical work on the
metal-poor populations of the halo \citep[e.g.,][]{Fr87}. The axis
ratio of the stellar distribution can be computed from
\begin{equation}
q = \left[ 1 + {\rho_1 \over \rho_0} \right]^{-1/\gamma}
\end{equation}
which gives $q = 0.983$, in other words, very round.  This
distribution function is not unique, as there are undoubtedly more
complicated multi-component models. Nonetheless, it is the simplest
distribution function that is consistent with the kinematical data. It
is interesting that the triaxial kinematics are consistent with a
near-spherical stellar density law.

\begin{figure}
\plotone{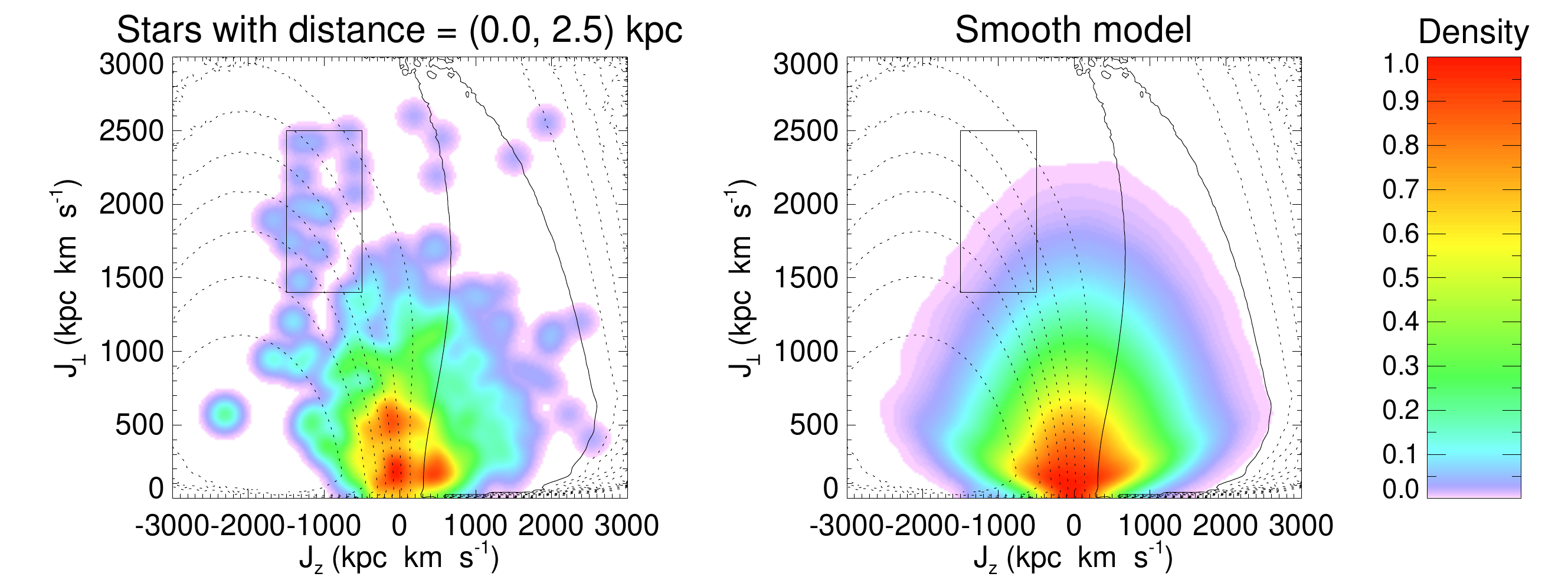}
\plotone{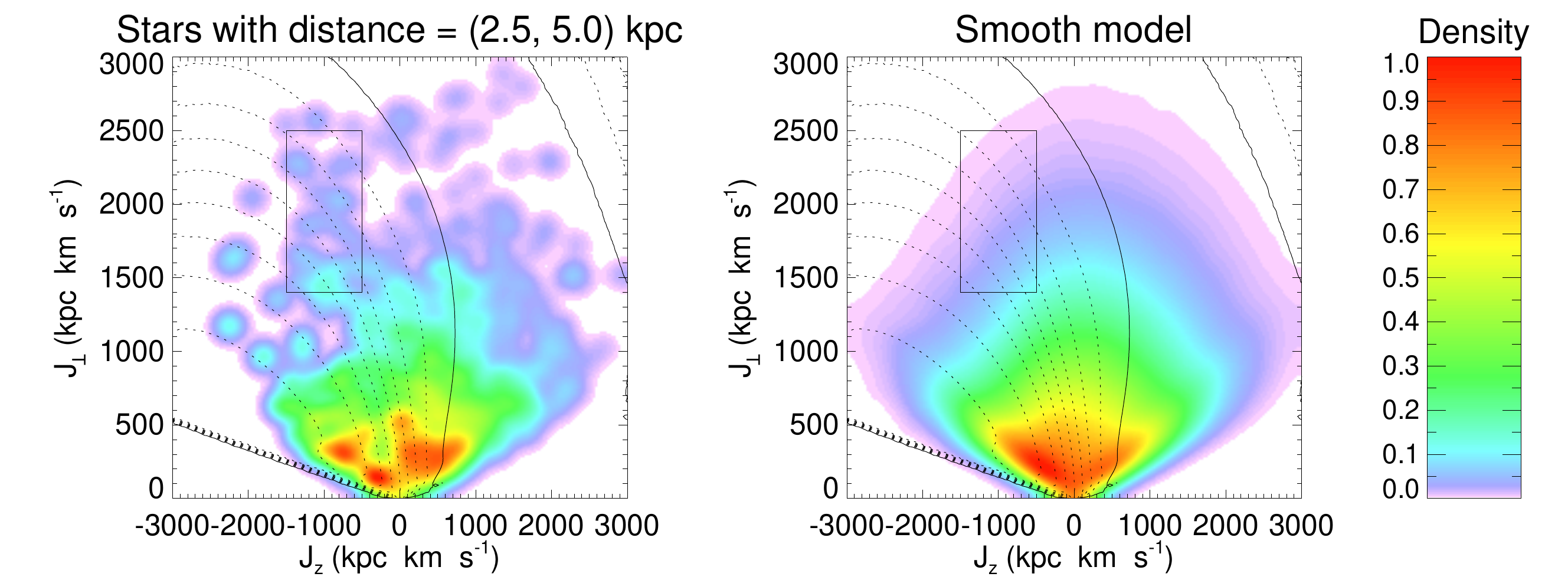}
\caption{The distribution of angular momentum for our subdwarfs (left)
  and for an artificial smooth halo (right). These plots cover the
  distance range 0 $-$ 2.5 kpc (top) and 2.5 $-$ 5 kpc (bottom).
  The contours denote our detection efficiency in steps of 10 per
  cent, where the solid line corresponds to the 90 per cent contour. The
  box corresponds to the location of the kinematic stream identified
  by \citet{He99}. In this coordinate system the Sun would lie at
  approximately $(-1800,\,0)\,\jun$.}
\label{fig:j_smooth}
\end{figure}

\begin{figure}
\plotone{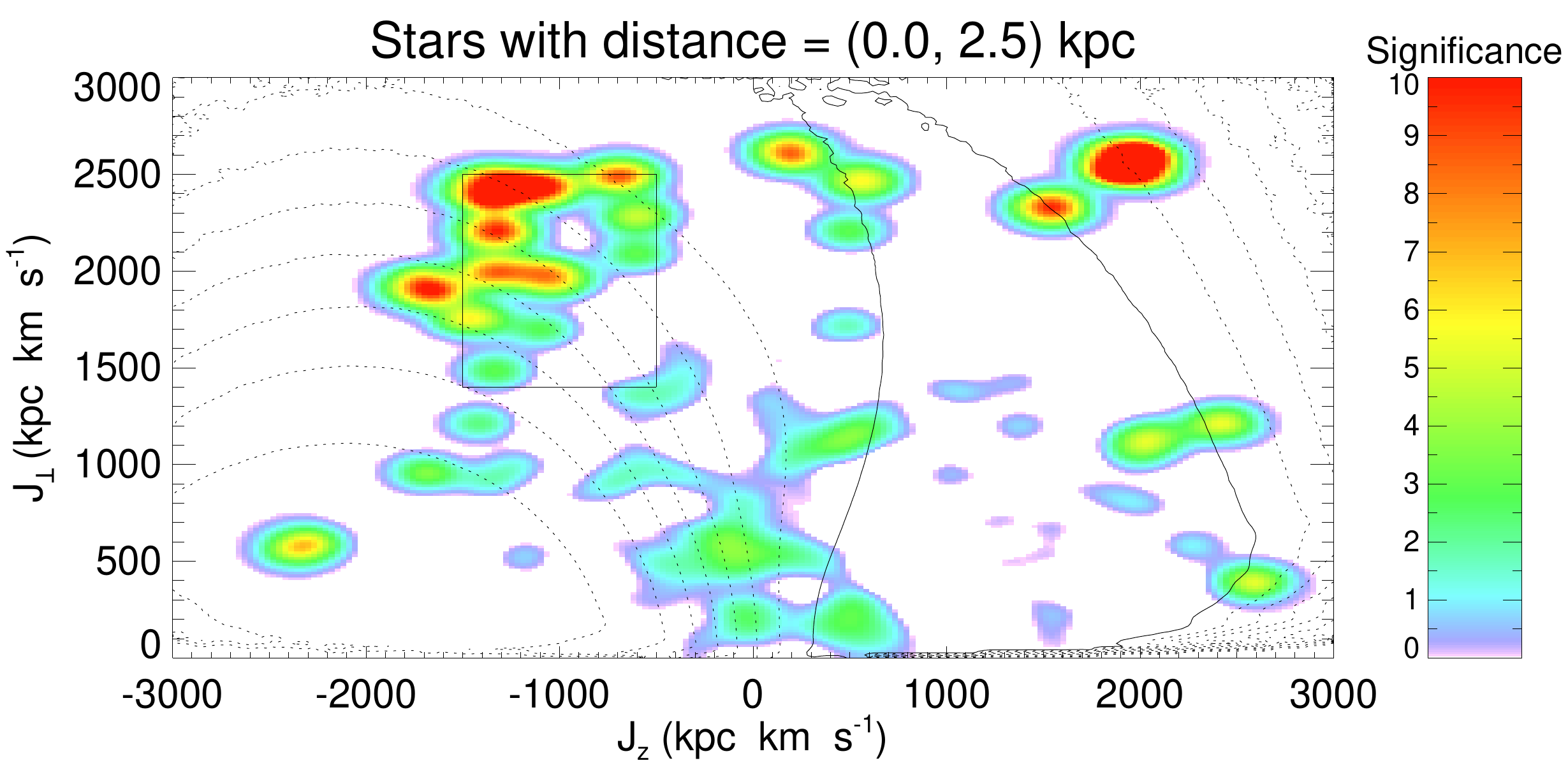}
\plotone{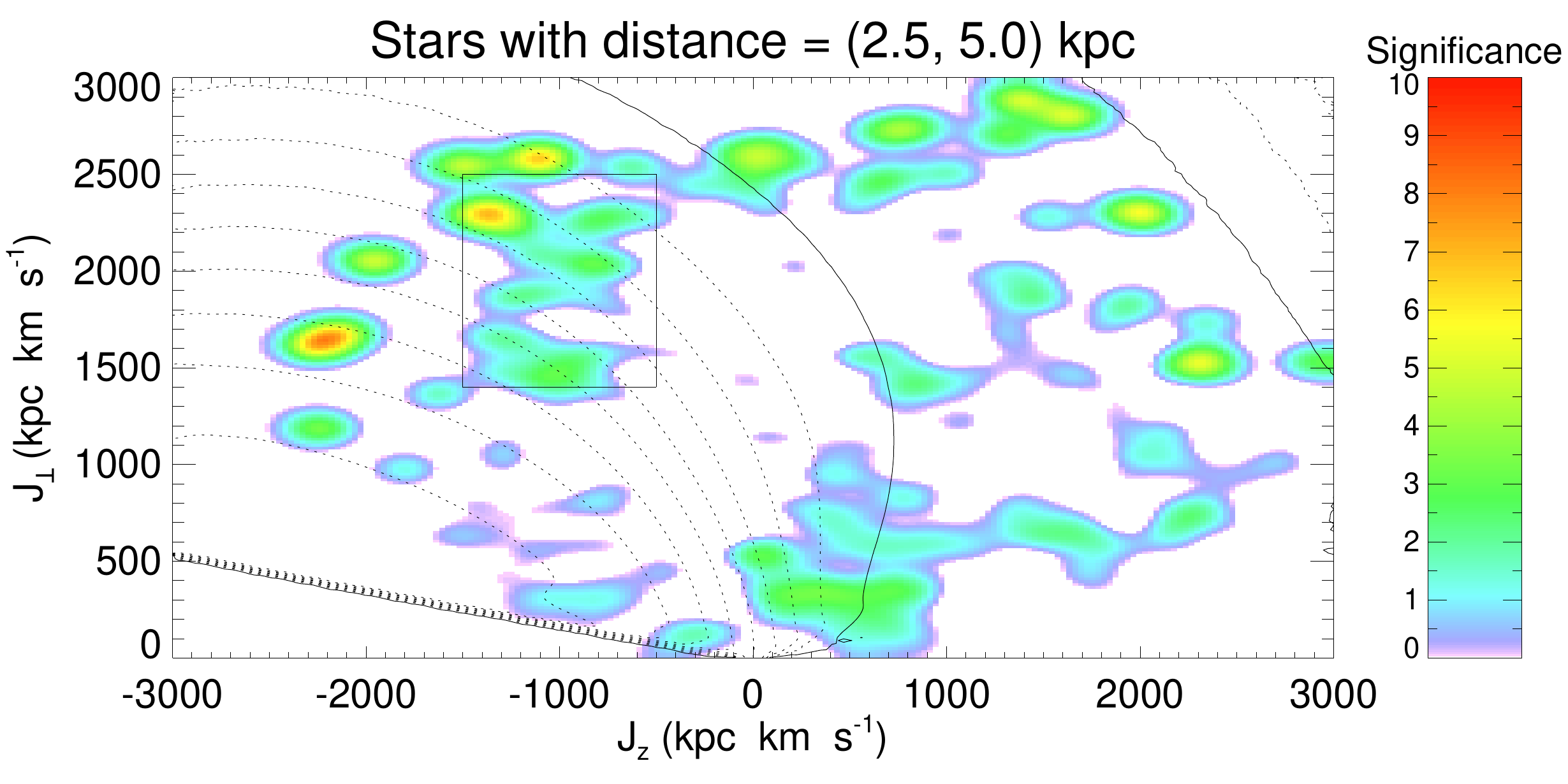}
\caption{The distribution of angular momentum residuals for subdwarfs
  in the distance range 0 $-$ 2.5 kpc (top) and 2.5 $-$ 5 kpc
  (bottom). The residuals are obtained after subtracting a smooth
  model from the observed distribution.  The contours and H99 boxes
  are as in Fig. \ref{fig:j_smooth}. In this coordinate system the Sun
  would lie at approximately $(-1800,\,0)\,\jun$.}
\label{fig:j_resid}
\end{figure}

\begin{figure}
\plotone{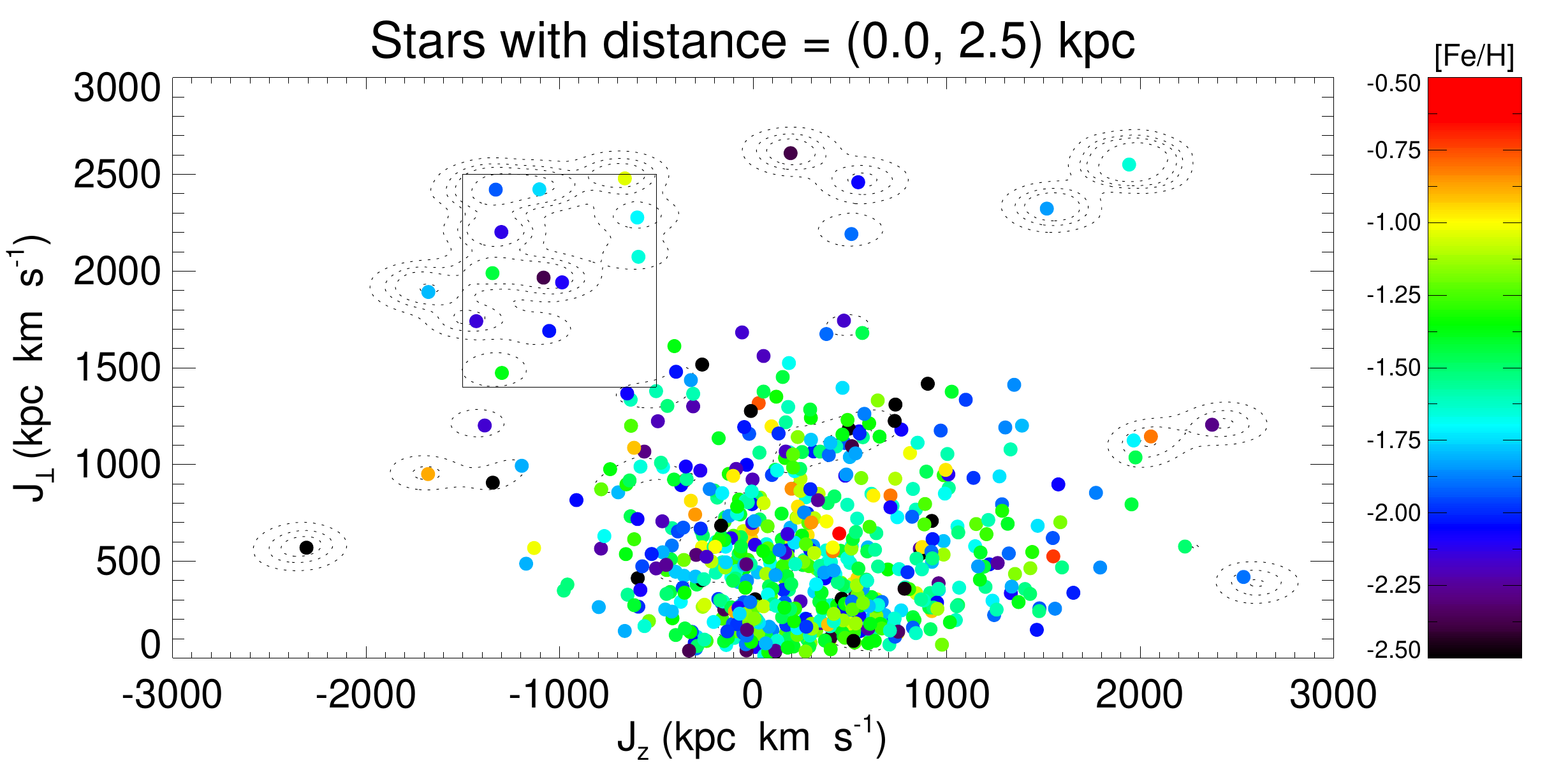}
\plotone{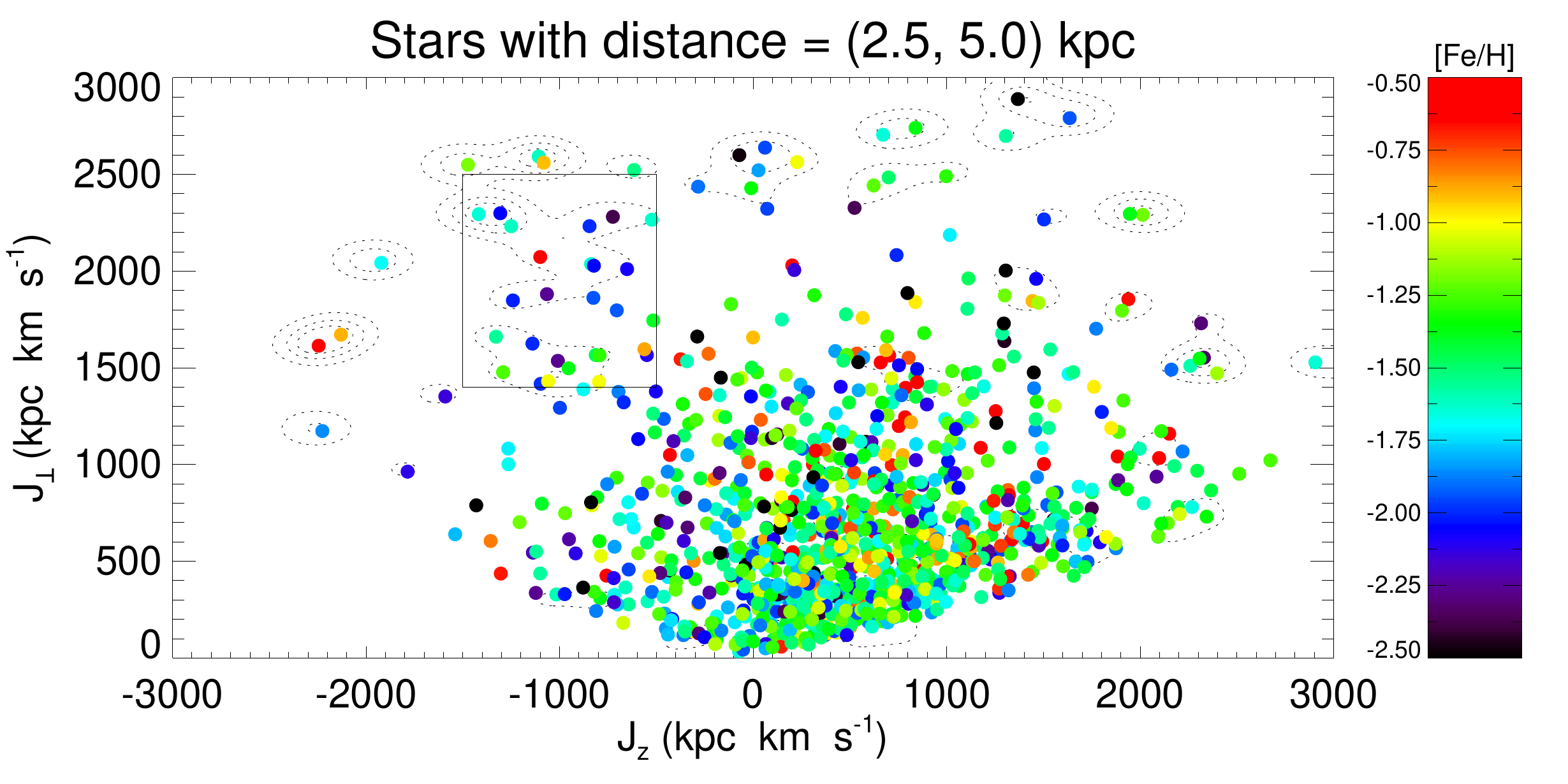}
\caption{The distribution of the angular momentum components of our
  subdwarfs in the distance range 0 $-$ 2.5 kpc (top) and 2.5 $-$ 5 kpc
  (bottom).  Stars are colour-coded according to $\feh$ and the
  colour-range saturates at -2.5 and -0.5.  Contours denote the
  residuals shown in Fig. \ref{fig:j_resid}, corresponding to 1, 3, 5
  and 10 sigma significance.  There are a number of stars which lie outside the
  range of this plot; these are given in Table \ref{tab:outliers}.
  In this coordinate system the Sun would lie at approximately $(-1800,\,0)\,\jun$.}
\label{fig:j_resid2}
\end{figure}

\section{Kinematic substructure}
\label{sec:substr}

\subsection{Quantification of Substructure}

The components of the angular momentum perpendicular and parallel to
the symmetry axis of the Galactic plane are
\begin{equation}
  \jperp = \left[(y\vz\!-\!z\vy)^2 + (z\vx\!-\!x\vz)^2\right]^{1\over2},\qquad\qquad J_z = \rx \vy - \ry \vx.
\end{equation}
For a solar neighbourhood sample, these two angular momentum
components are essentially $\jz \sim -{\rm r}_\odot \vy$ and $\jperp
\sim {\rm r}_\odot \left| \vz \right|$.  Our median errors on $(\jz,
\jperp)$ are (410, 255) ${\rm \,km^2\,s^{-1}}$. If we restrict our
sample to the $645$ subdwarfs within 2.5 kpc, then the errors are
reduced by a factor of $\sim 35$ per cent.  Use of these coordinates
is a common approach~\citep[e.g. H99,][]{Ch00}, as they are adiabatic
invariants in a spherical potential.  Note that, unlike H99, we use a
right-handed coordinate system, so our values of $\jz$ take an
opposite sign to theirs. This means that in our coordinate system the
Sun would lie at approximately $(-1800,\,0)\,\jun$.
We weight each star according to the inverse
of its detection efficiency to account for kinematic bias and smooth
the distribution using a Gaussian kernel with FWHM of $(300,150)\jun$,
which is the same order of magnitude as the errors. The resulting
distribution is plotted in Fig. \ref{fig:j_smooth}.


To test for substructure, we generate an artificial sample of halo
stars using the observed distances and directions of stars in our
sample, but generate velocities according to the trivariate Gaussian
with means and dispersions given in Table \ref{tab:uvw}. We smooth
this distribution and subtract it from the smoothed observed
distribution (see Fig. \ref{fig:j_smooth}). The resulting residual
plot is shown in Fig. \ref{fig:j_resid}, where the colour-scale shows
the significance of overdensities. The significance is quantified by
generating a series of efficiency-corrected realisations from our
distributions, all of which have the same number of stars as the
observed sample. The scatter in these realisations provides an
estimate of $\sigma(\jz,\jperp)$, from which we can deduce the
significance of our overdensities. We show the overdensities for two
distance ranges (the 645 stars with $\D<2.5$ kpc and the 1072 with
$2.5<\D<5$ kpc) and include contours showing the detection efficiency.

Some of the apparent overdensities in the outer regions are artefacts.
The smooth model predicts very few stars in the outer regions and
hence $\sigma(\jz,\jperp)$ is close to zero, which exaggerates the
significance of individual outliers. For example, the large clump
around $(\jz,\jperp)=(2000,2500)\jun$ in the upper panel of
Fig. \ref{fig:j_resid} is actually composed of only one star.  This
can be seen on comparison with Fig. \ref{fig:j_resid2}, which shows
the unbinned ($\jz,\jperp$) distribution for each of the stars.  Note
that the figures confirm that our sample is free from significant disc
contamination. If there were significant thin or thick disc stars,
they would be visible in the region around $-2500 < J_{\rm z} < -1500
\jun$ and $J_\perp < 600 \jun$.

\subsection{The H99 Kinematic Stream}
\label{sec:h99}

The most noticeable feature in Fig. \ref{fig:j_resid2} for stars
within 2.5 kpc is the asymmetry in $\jz$ for stars with $1500 < \jperp
< 2500 \jun$. There is a significant number of stars located around
$(\jz,\jperp) \approx (-1000,2000)\jun$. This corresponds to the solar
neighbourhood kinematic stream first identified by \citet{He99}. They
analysed a sample of nearby (i.e. $<$ 1 kpc) halo stars with full
six-dimensional phase space information and found that $\sim$ 10 per
cent come from a single coherent 
structure. Subsequently, \citet[][hereafter K07]{Ke07} probed a
larger volume ($D<2.5$ kpc) and found a smaller fraction ($\sim 5$ per
cent) of stars belonging to this stream. Restricting their sample to 1
kpc, they find a larger fraction ($\sim 9$ per cent) in agreement with
H99, although this is not surprising since there is significant
overlap between the H99 and K07 samples.  \citet{Ch00} also studied an
extended version of the H99 halo sample and found a smaller fraction
than H99, although the exact percentage is unclear. K07 defined the
location of the H99 region as $-1500 < \jz < -500$ and $1400 < \jperp
< 2500$, which is indicated in Figs. \ref{fig:j_smooth} $-$ \ref{fig:j_resid2}.

There has also been related work on streams in the solar neighbourhood
subdwarfs by \citet{Go03b}. This analysis, which was based on a large
sample of objects with proper motion measurements but without radial
velocities or distance determinations, used a Bayesian likelihood
analysis to investigate various properties of the nearby subdwarf
population.  Instead of looking for direct evidence of cold kinematic
streams, he quantified the amount of granularity in the velocity space
and concluded that no more than 5 per cent of local halo stars (within
$\sim 300$ pc) can come from a single cold stream.  Although this
number appears to be in conflict with the original results from H99,
there is actually no discrepancy as the H99 stream is formed from two
separate clumps in velocity space.

The H99 stream is clearly visible in our data, as can be seen in the
top panel of Fig. \ref{fig:j_resid}, although it appears more diffuse
than the detection in K07 (c.f., their fig. 11). Note that this
difference cannot be explained by the size of our observational
errors, since the median error on $(\jz, \jperp)$ for our H99 stars is
(240, 209) ${\rm \,km^2\,s^{-1}}$.  Closer inspection of the upper
panel of Fig. \ref{fig:j_resid2} shows that the H99 box contains two stars near
the centre along with a more diffuse component of 10 stars. These two
stars lie almost exactly in the centre of the box at $(\jz,\jperp)
\approx (-1035, 1954)\jun$. The mean metallicity of the two stars is
-2.2 dex, which is more metal poor than our full subdwarf
population (-1.5 dex). However, the mean metallicity of all 12 stars
in the H99 region is -1.8 dex with a dispersion of 0.4 dex, which is
comparable to the overall distribution.

\begin{figure}
\plotone{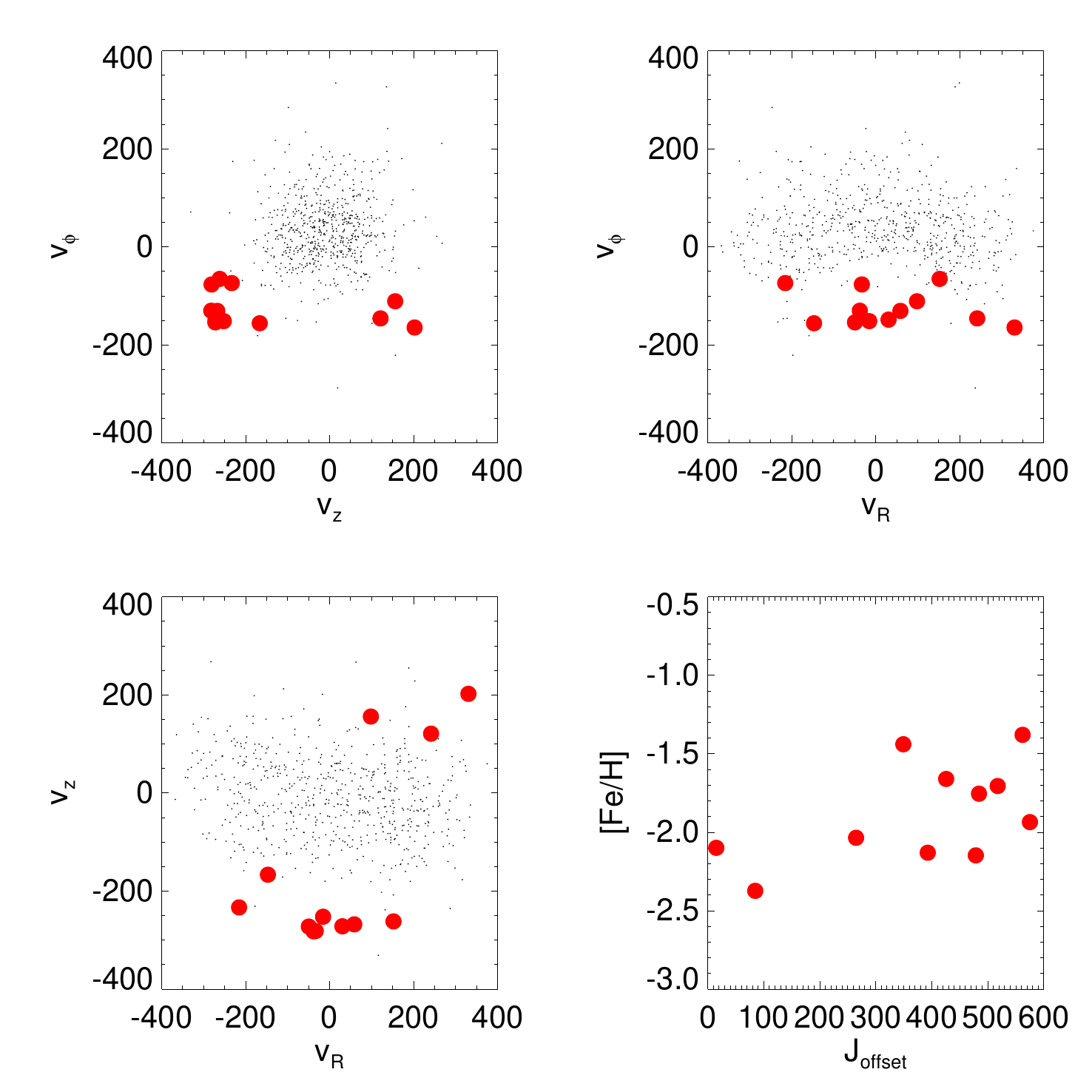}
\caption{Properties of our H99 stars (see Section
  \ref{sec:h99}). Upper panels and lower-left panel show the velocity
  distributions for all stars within 2.5 kpc (small dots) and stars in
  the H99 region (large circles). The bottom-right panel shows $\feh$
  as a function of offset from the centre of the H99 box in the 2D
  $(\jz,\jperp)$ plane, which lies at $(-1000,1950)\jun$.}
\label{fig:h99}
\end{figure}

The stars in and around the H99 region are illustrated in
Fig. \ref{fig:h99}, which shows their velocities and offset in the
($\jz,\jperp$) plane from the centre of the H99 box (the mean error in
${\rm J_{offset}}$ is $196\jun$). Given that a smooth halo
distribution should give $\sim 4$ stars in this region, we suspect
that not all of these 12 stars are associated with the H99 feature.
The distance distribution of these stars is approximately uniform
between $\sim1$ and $\sim2.5$ kpc, with no obvious trend towards
smaller distances (unlike K07, who found 67 per cent of their H99
stars were within 1 kpc). However, we note that our sample is biased
against $\D\la1$ kpc, as can be seen from Fig. \ref{fig:non_kin},
which implies that we could be missing stars from this feature.  If we
consider the more distant stars with $\D>2.5$ kpc as shown in the
lower panel of Fig. \ref{fig:j_resid2}, we find further potential
members of the H99 region. However, these stars appear to come from a
more homogeneous background of stars with high angular momentum.

We wish to quantify the prevalence of the H99 stream. We first
need to correct for the detection efficiency in order to account for
the kinematic bias in our sample. However, the efficiency in
Section \ref{sec:bias} is not directly applicable for stars in the H99
stream, as they do not have the same velocity distribution as the
underlying smooth halo. To account for this we re-calculate the
efficiency for stars with H99 kinematics. Using this efficiency we
predict that up to $\sim 7$ per cent of all halo stars within 2.5 kpc lie
in the H99 box, which is significantly larger than the fraction
expected from an entirely smooth halo ($\sim 1$ per cent).

\begin{table*}
\begin{tabular}{lccccccc}
\hline
SDSS ID & $\jz$ & $\jperp$ & $\vr$ & $\vphi$ & $\vth$ & Distance & $\feh$\\
&$(\jun)$ & $(\jun)$ & $(\kms)$ & $(\kms)$ & $(\kms)$ & (kpc) & (dex)\\
\hline
SDSS J002012.32+003954.3 &    3368.9 $\pm$     909.9 &    1597.8 $\pm$     560.9 &     365.8 $\pm$      64.7 &     377.0 $\pm$      98.0 &      21.0 $\pm$      55.1 &  4.80 $\pm$  0.52 & -1.15 $\pm$  0.12\\
SDSS J003434.00-011244.6 &   -3779.3 $\pm$     516.5 &    1855.1 $\pm$     365.3 &     -82.2 $\pm$      50.1 &    -433.8 $\pm$      57.7 &    -139.6 $\pm$      37.8 &  3.43 $\pm$  0.33 & -1.88 $\pm$  0.13\\
SDSS J003948.05+003701.5 &    -135.4 $\pm$     541.0 &    3119.4 $\pm$     449.6 &      21.4 $\pm$      53.3 &     -15.3 $\pm$      61.1 &    -333.8 $\pm$      46.4 &  3.43 $\pm$  0.32 & -2.56 $\pm$  0.15\\
SDSS J010450.81-011244.8 &   -1162.1 $\pm$     270.3 &    3568.3 $\pm$     303.5 &    -151.1 $\pm$      33.5 &    -136.3 $\pm$      32.2 &     410.8 $\pm$      31.3 &  1.76 $\pm$  0.20 & -1.38 $\pm$  0.11\\
SDSS J010643.56+005744.2 &     443.5 $\pm$     801.9 &    3221.2 $\pm$     812.0 &     164.0 $\pm$      63.4 &      45.6 $\pm$      82.0 &    -301.4 $\pm$      70.9 &  4.99 $\pm$  0.57 & -0.78 $\pm$  0.17\\
SDSS J010747.56-010447.0 &   -3240.1 $\pm$     557.3 &    2043.0 $\pm$     396.2 &    -124.9 $\pm$      51.3 &    -356.5 $\pm$      59.6 &    -180.1 $\pm$      42.7 &  3.42 $\pm$  0.32 & -2.18 $\pm$  0.14\\
SDSS J013935.21-002545.9 &    -773.7 $\pm$     420.3 &    3233.2 $\pm$     428.1 &      94.3 $\pm$      35.4 &     -82.2 $\pm$      44.9 &    -327.5 $\pm$      37.7 &  3.32 $\pm$  0.38 & -1.44 $\pm$  0.11\\
SDSS J015712.77+011137.1 &    -929.6 $\pm$     273.1 &    3748.4 $\pm$     375.6 &    -112.3 $\pm$      28.1 &    -103.4 $\pm$      31.0 &     409.3 $\pm$      35.5 &  2.01 $\pm$  0.22 & -1.35 $\pm$  0.12\\
SDSS J015755.12+000416.1 &     661.4 $\pm$     566.9 &    4171.8 $\pm$     515.7 &      -4.6 $\pm$      39.2 &      68.9 $\pm$      58.1 &    -416.5 $\pm$      43.5 &  3.30 $\pm$  0.41 & -1.43 $\pm$  0.12\\
SDSS J020359.08-005927.9 &    -742.4 $\pm$     588.5 &    3857.5 $\pm$     641.2 &     -51.4 $\pm$      38.0 &     -73.1 $\pm$      58.1 &     355.9 $\pm$      52.6 &  4.35 $\pm$  0.45 & -1.74 $\pm$  0.16\\
SDSS J021724.92+003127.3 &    -904.2 $\pm$     541.7 &    5078.4 $\pm$     704.6 &      56.8 $\pm$      39.6 &     -89.4 $\pm$      53.8 &     -478.2 $\pm$      55.9 &  3.86 $\pm$  0.42 & -1.97 $\pm$  0.12\\
SDSS J030816.92+011513.3 &    1956.8 $\pm$     854.4 &    3322.5 $\pm$     605.5 &     113.4 $\pm$      29.8 &     187.7 $\pm$      77.7 &     306.3 $\pm$      49.8 &  3.53 $\pm$  0.46 & -1.45 $\pm$  0.14\\
\hline
\end{tabular}
\caption{Retrograde and Prograde Outliers}.
\label{tab:outliers}
\end{table*}

\subsection{Retrograde and Prograde Outliers}
\label{sec:outliers}

Various authors have noted the presence of prograde or retrograde
outliers when analysing distributions of halo velocities. For example,
\citet{Ve04} identified halo stars on extreme retrograde orbits
($\vphi > 200 \kms$) and found them to be significantly more metal
poor than typical halo stars. K07 similarly saw an excess of
retrograde stars in their halo sample and noted a deficiency in $\feh$
(they found 6 stars with $\langle\feh\rangle = -2.1$ dex). K07 argued
that these stars probably belong to a tidally disrupted stream. In
addition, K07 found 3 metal-poor stars on extreme prograde orbits,
although they postulated that these were more consistent with the
underlying halo distribution.

In our sample, we find many stars on extreme orbits, as can be seen
from Fig. \ref{fig:j_resid2}. There are also a number of stars whose
angular momenta is sufficiently large for them to lie outside the
region of this plot; those stars are listed in Table
\ref{tab:outliers}.

However, none of the low $\jperp$ outliers in Fig \ref{fig:j_resid2}
look particularly clumped and neither the prograde nor retrograde
outliers are significantly in excess of expectations of a smooth
distribution. If we analyse stars with $\jperp < 2000$, then we find 5
stars with $\jz < -2000 \jun$ and 31 stars with $\jz > 2000 \jun$. Our
smooth model predicts 2.4 and 30.3 stars in these regions,
respectively. Furthermore, neither of these two samples are
particularly metal-poor; the mean metallicity is -1.6 dex and -1.4 dex
for the prograde and retrograde outliers, respectively.  Note that the
strong asymmetry in the number of outliers is due to the fact that our
detection efficiency is significantly lower for $\jz<0$.

\begin{table*}
\begin{tabular}{ccccccccc}
\hline
\sko & SDSS ID & $\jz$ & $\jperp$ & $\vr$ & $\vphi$ & $\vth$ & Distance & $\feh$\\
&&$(\jun)$ & $(\jun)$ & $(\kms)$ & $(\kms)$ & $(\kms)$ & (kpc) & (dex)\\
\hline

a & SDSS J215939.43-004835.6 &     542.9 $\pm$     347.4 &    2459.5 $\pm$     262.2 &     142.4 $\pm$      73.7 &      71.0 $\pm$      48.6 &     320.4 $\pm$      57.2 &  0.94 $\pm$  0.09 & -2.07 $\pm$  0.13\\ 
a & SDSS J004328.45-001700.0 &     507.9 $\pm$     700.8 &    2191.7 $\pm$     353.6 &     162.8 $\pm$      46.4 &      60.8 $\pm$      71.6 &    -258.5 $\pm$      48.9 &  1.55 $\pm$  0.16 & -1.90 $\pm$  0.12\\ 
a & SDSS J014424.63+003442.0 &     193.6 $\pm$     666.6 &    2609.8 $\pm$     301.1 &     135.0 $\pm$      26.8 &      21.7 $\pm$      60.8 &    -286.9 $\pm$      44.8 &  2.06 $\pm$  0.19 & -2.38 $\pm$  0.13\\ 

a$^\star$ & SDSS J012713.03+011341.8 &     452.4 $\pm$     850.1 &    2529.1 $\pm$     375.0 &      46.8 $\pm$      33.6 &      53.2 $\pm$      74.3 &     294.7 $\pm$      50.7 &  1.29 $\pm$  0.13 & -1.90 $\pm$  0.25\\ 
a$^\star$ & SDSS J014332.60-010726.7 &     397.6 $\pm$     711.3 &    2281.7 $\pm$     344.3 &    -212.0 $\pm$      25.6 &      44.9 $\pm$      65.1 &    -252.4 $\pm$      39.7 &  2.01 $\pm$  0.19 & -1.76 $\pm$  0.13\\ 
a$^\star$ & SDSS J030435.92-002403.2 &      51.3 $\pm$     770.0 &    2233.4 $\pm$     631.6 &    -217.6 $\pm$      35.4 &       5.9 $\pm$      68.9 &    -255.1 $\pm$      56.5 &  1.08 $\pm$  0.10 & -2.06 $\pm$  0.11\\ 

\\

b & SDSS J031005.44-001459.4 &     671.6 $\pm$     797.2 &    2704.8 $\pm$     774.1 &     181.8 $\pm$      33.4 &      61.3 $\pm$      71.3 &    -236.3 $\pm$      61.7 &  4.36 $\pm$  0.52 & -1.65 $\pm$  0.12\\ 
b & SDSS J031049.12+005059.3 &     997.5 $\pm$     712.7 &    2490.8 $\pm$     614.4 &     -25.3 $\pm$      28.4 &      90.8 $\pm$      62.9 &    -216.5 $\pm$      49.7 &  4.33 $\pm$  0.50 & -1.30 $\pm$  0.11\\ 
b & SDSS J031123.47+001249.6 &    1305.7 $\pm$     852.2 &    2698.5 $\pm$     632.5 &    -189.8 $\pm$      27.2 &     114.6 $\pm$      71.5 &    -223.1 $\pm$      49.2 &  4.95 $\pm$  0.63 & -1.57 $\pm$  0.11\\ 
b$^\dagger$ & SDSS J030641.38-002338.0 &    1221.2 $\pm$    1285.3 &    3248.5 $\pm$    1198.1 &     -63.2 $\pm$      42.8 &      99.9 $\pm$     102.0 &    -245.5 $\pm$      86.1 &  6.31 $\pm$  0.85 & -1.17 $\pm$  0.11\\ 
b$^\dagger$ & SDSS J030948.29-000111.3 &    1371.6 $\pm$    1650.7 &    2433.0 $\pm$    1455.8 &      15.6 $\pm$      49.7 &     102.7 $\pm$     121.0 &    -162.2 $\pm$     108.3 &  7.87 $\pm$  0.95 & -1.61 $\pm$  0.12\\ 
b$^\dagger$ & SDSS J031120.73-000329.4 &    1655.1 $\pm$    1097.4 &    2735.3 $\pm$    1050.5 &     245.6 $\pm$      37.1 &     141.2 $\pm$      89.6 &    -216.2 $\pm$      76.9 &  5.45 $\pm$  0.70 & -1.10 $\pm$  0.16\\ 

\\

c & SDSS J015840.78-000115.6 &    1305.3 $\pm$     690.3 &    2002.9 $\pm$     461.6 &    -177.6 $\pm$      38.9 &     130.4 $\pm$      67.5 &    -183.8 $\pm$      47.4 &  4.12 $\pm$  0.40 & -2.61 $\pm$  0.17\\ 
c & SDSS J015840.90+002900.8 &    1367.9 $\pm$     864.8 &    2889.3 $\pm$     729.4 &      87.6 $\pm$      49.5 &     133.1 $\pm$      82.3 &    -258.5 $\pm$      62.3 &  4.60 $\pm$  0.52 & -2.61 $\pm$  0.18\\ 
c$^\ddagger$ & SDSS J015806.00+003419.5 &    1825.3 $\pm$    1378.6 &    2688.8 $\pm$    1247.4 &     228.8 $\pm$      62.3 &     162.2 $\pm$     119.1 &    -201.9 $\pm$      97.3 &  6.55 $\pm$  0.82 & -2.51 $\pm$  0.12\\ 

\hline
\end{tabular}
\caption{Stars which are candidate members of the three new kinematic
  overdensities (\sko a-c). Note that the errors on the velocities and
  angular momenta can be highly correlated; they are included here
  to give an indication of the relative uncertainties. Notes:
  $^\star$ The Sloan Stellar Parameter Pipeline flagged the parameters
  for these stars as `cautionary' due to issues regarding the
  H$\alpha$ line strength; $^\dagger$ The distances for these 
  stars are greater than 5 kpc, which means that the derived
  velocities are subject to large uncertainties; $^\ddagger$ The Sloan 
  Stellar Parameter Pipeline flagged the parameters for this star as
  `cautionary' due to the presence of a strong $G$-band
  feature in the spectrum, and the distance is greater than 5 kpc and
  so the velocities are uncertain. }
\label{tab:sko}
\end{table*}

\section{New Kinematic Overdensities}

\label{sec:new}

There are a number of other possible clumps in our sample. We discuss
three of them, which we label Sloan Kinematic Overdensities (\sko), in
the following section. The associated stars are given in Table
\ref{tab:sko}.

\subsection{\sko a}
\label{sec:skoa}

One interesting feature in the nearby sample shown in the upper-panel
of Fig. \ref{fig:j_resid2} is the collection of 3 retrograde stars
centred on $(\jz,\jperp)\approx(410,2420)\jun$. The three stars
are all more metal-poor than the halo average, with
$\langle\feh\rangle$ of -2.1 dex, and are not clumped spatially (see
Table \ref{tab:sko}). The stars are enclosed by a circular region of
the ($\jz,\jperp$) plane centred on (414.8, 2420.3) with radius 350
$\jun$. Our smooth model predicts that there should be only 0.1 per
cent of the sample in this region, which is significantly smaller than
the observed fraction of 0.5 per cent. We shall refer to this
potential overdensity as \sko a.

When constructing our subdwarf sample, we enforced a strict cut on the
quality flags that are raised by the SSPP (see Section
\ref{sec:spectra}). The SSPP includes two categories of flags:
critical and cautionary. For the latter category, it is often possible
to determine the stellar parameters, although they should be treated
with caution. The vast majority of these stars are flagged because the
SSPP pipeline raised issues regarding the H$\alpha$ line strength
\citep{Le08a}. In particular, this flag marks stars where `there
exists a strong mismatch between the strength of the predicted
H$\alpha$ line index, based on the measured H$\delta$ line index'.  If
we include stars with cautionary flags into our subdwarf sample, then
we gain an additional $\sim700$ stars, of which three appear to be
associated with the \sko a feature.  All of these stars are have
similar metallicities to the \sko a members mentioned above (see Table
\ref{tab:sko}).

\begin{figure}
\begin{center}
\includegraphics[width=\hsize]{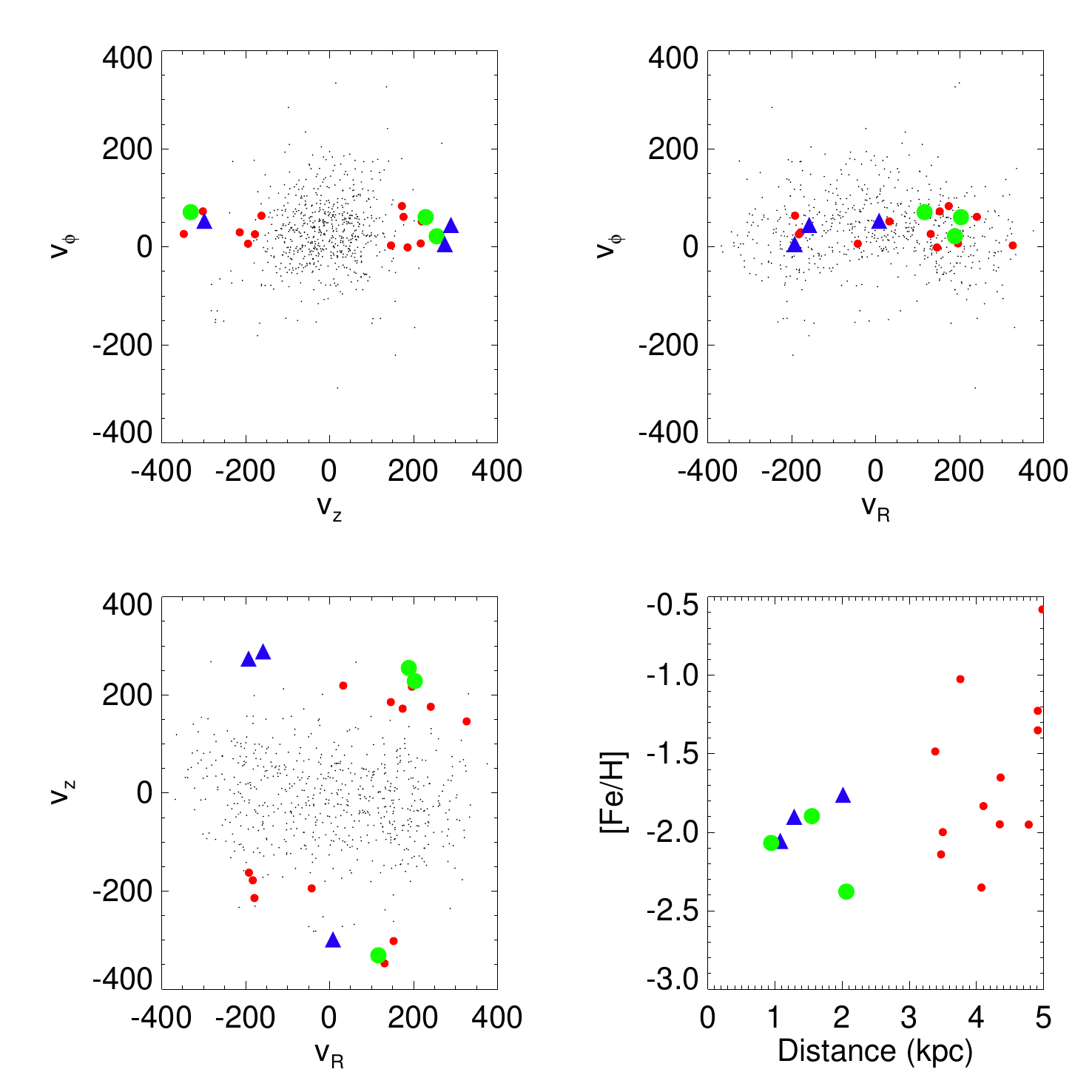} 
\end{center}
\caption{Properties of the \sko a overdensity (see Section
  \ref{sec:skoa} and Table \ref{tab:sko}). The upper two panels and
  lower left panel show the velocity distributions for all stars
  within 2.5 kpc of the Sun (small points), members of the \sko a
  overdensity (large circles), additional members with cautionary SSPP
  flags (triangles), and potential members with distance greater than
  2.5 kpc (small circles). The members with $\D < 2.5$ kpc are
  enclosed within a 350 $\jun$ circle centred on (413.6, 2419.9)
  $\jun$, while the potential members with $2.5 < \D < 5$ kpc are
  enclosed within a larger radius (500 $\jun$). The bottom right panel
  shows $\feh$ as a function of heliocentric distance for stars with
  offset in the 2D $(\jz,\jperp)$ plane less than 500 $\jun$ from the
  centre of the \sko a clump.}
\label{fig:skoa}
\end{figure}

We plot the velocities of the six \sko a stars in Fig. \ref{fig:skoa},
along with the distribution of $\feh$ as a function of heliocentric
distance. Also included in this figure are the stars in a looser
selection region with radius 500 $\jun$ and with distances up to 5
kpc. It seems that a number of the more distant stars may also be
associated with this overdensity.

\begin{figure*}
\plotone{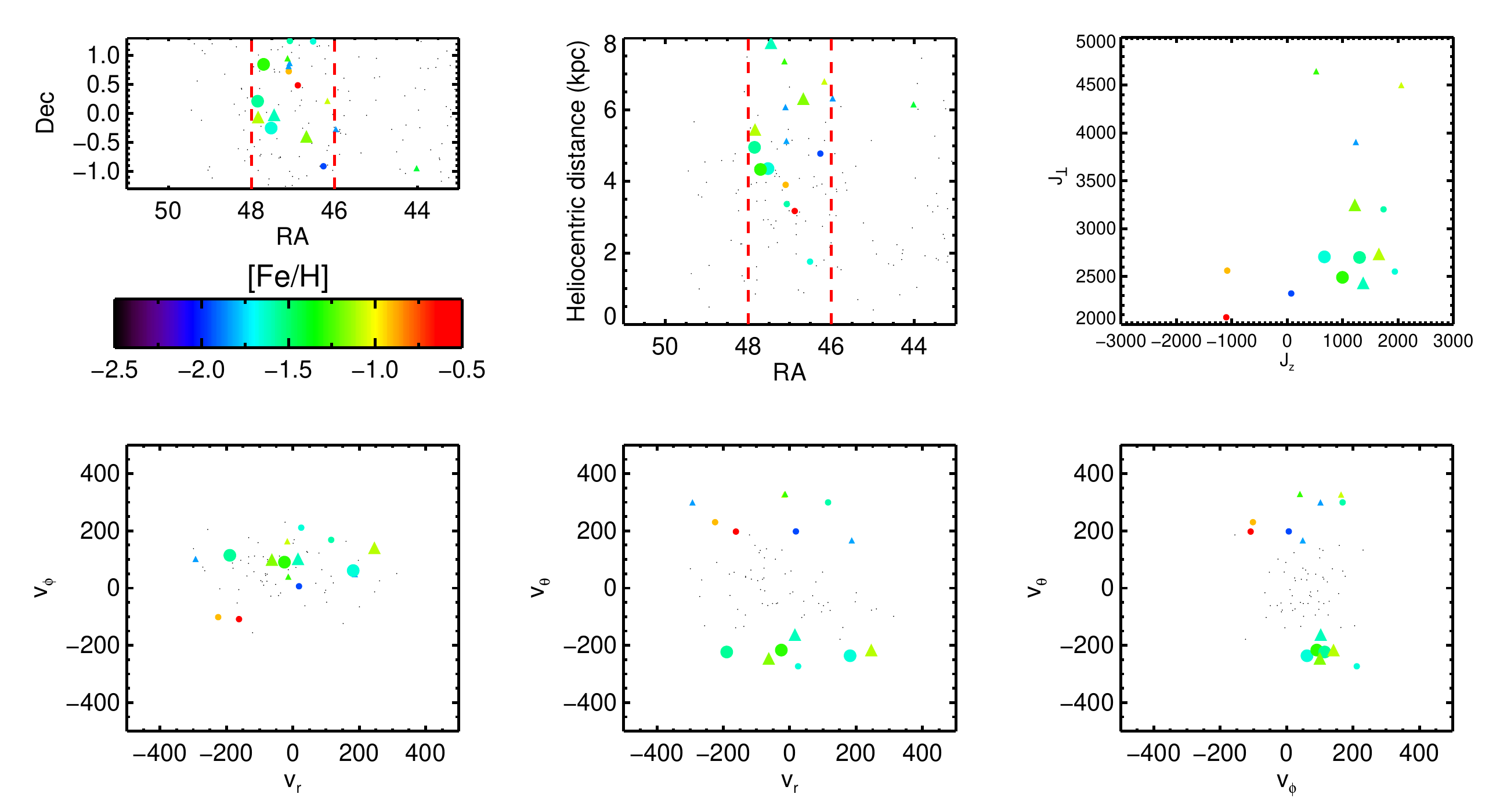}
\caption{Properties of the \sko b overdensity (see Section
  \ref{sec:skobc} and Table \ref{tab:sko}).
The coloured symbols
  denote stars with $\jperp>2000\jun$ and are colour-coded according
  to metallicity, with \sko b members (large symbols) differentiated
  from other high-$\jperp$ stars (small symbols). The triangles denote
  stars with $\D>5$ kpc, which have larger uncertainties in their
  kinematics. We also show stars with $\jperp<2000\jun$ for comparison
  (small points). The vertical dashed lines denote the right ascension
  range $46^\circ<\alpha<48^\circ$. Note the lack of high-$\jperp$
  stars outside of this range.}
\label{fig:skob}
\end{figure*}
\begin{figure*}
\plotone{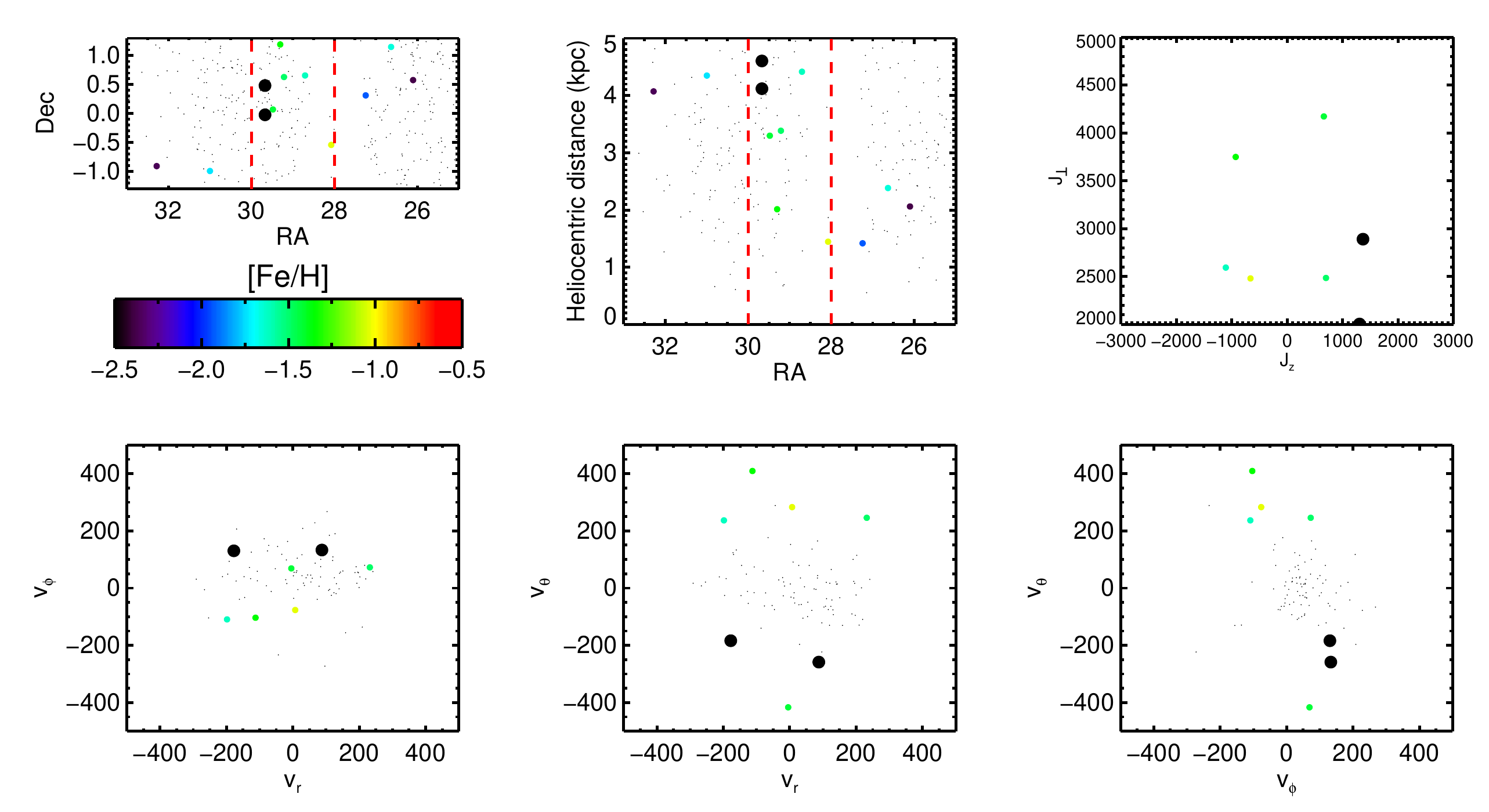}
\caption{Properties of \sko c sample stars in the range
$28^\circ<\alpha<30^\circ$ (see Section \ref{sec:skobc} and Table
\ref{tab:sko}). The upper-middle and upper-left panels also show
stars outside this $\alpha$ range for comparison.
The coloured symbols denote stars with $\jperp>2000\jun$ and are
colour-coded according to metallicity, with the two metal-poor \sko c
members (large symbols) differentiated from other high-$\jperp$ stars
(small symbols).}
\label{fig:skoc}
\end{figure*}

\begin{figure}
\plotone{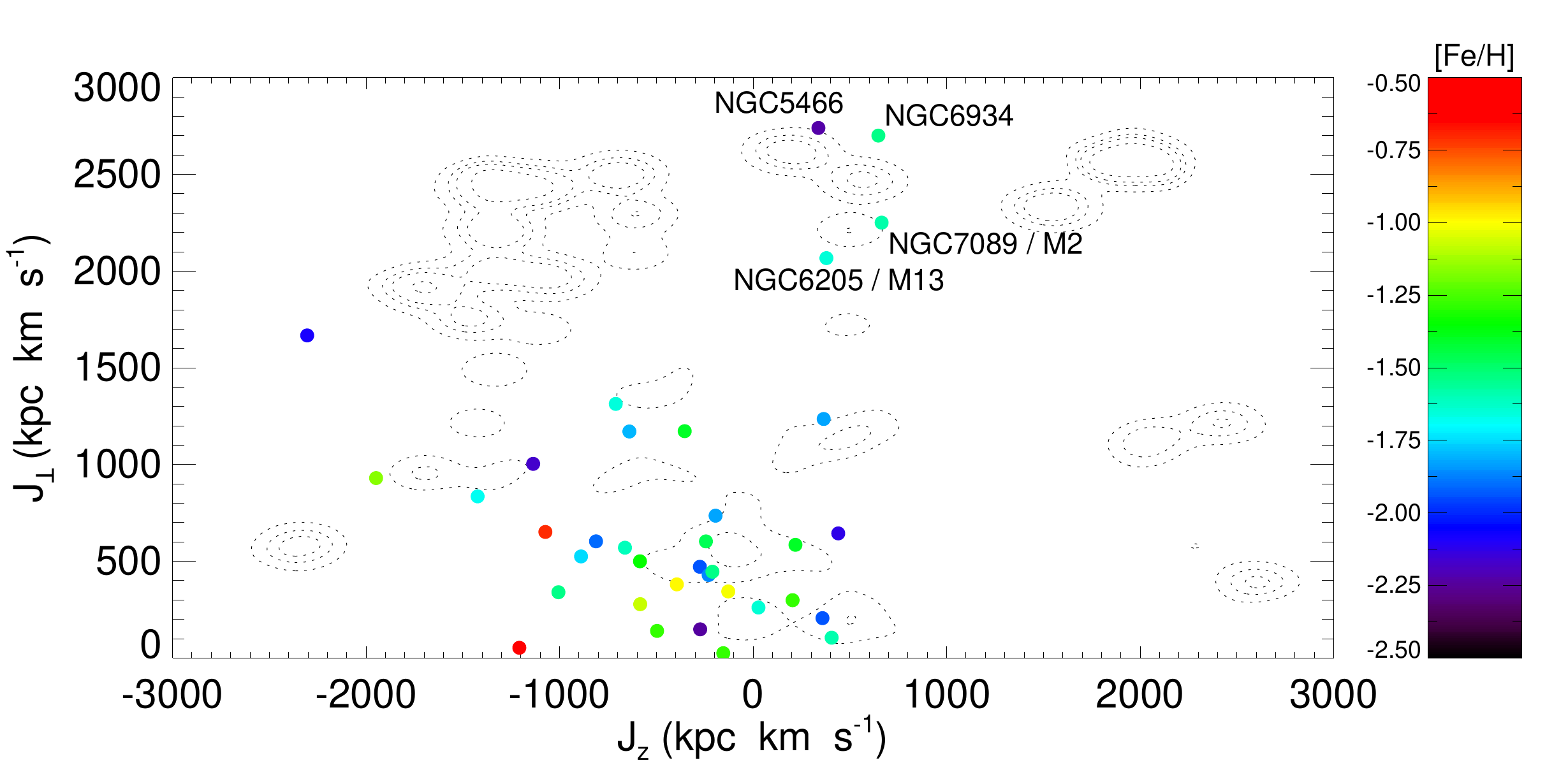}
\caption{Angular momentum distribution of the 38 globular clusters from
  \citet{Di99} with 6D phase-space, where the colour corresponds to cluster
  metallicity. Contours denote the kinematic overdensities for halo
  stars within 2.5 kpc (see Section \ref{sec:substr}), corresponding
  to 1, 3, 5 and 10 sigma significance. Note the four
  globular clusters potentially associated to the overdensity at 
  $(\jz,\jperp)\approx(500,2500)\jun$; these are NGC5466, NGC6934,
  NGC7089/M2, NGC6205/M13. In this coordinate system the Sun would lie
  at approximately $(-1800,\,0)\,\jun$.}
\label{fig:j_gc}
\end{figure}

Although the above discussion of \sko a appears tentative, there is
striking evidence supporting an accretion scenario for these stars. In
Fig. \ref{fig:j_gc}, we overplot the locations of globular clusters
using data from \citet{Di99} on our distribution of kinematic
overdensities in the $(\jz,\jperp)$ plane. We immediately see the
clump of four clusters standing out around \sko a. These clusters are
NGC5466, NGC6934, NGC7089/M2 and NGC6205/M13, which have metallicity
-2.22, -1.54, -1.58 and -1.65 dex, respectively. Of these four
clusters, NGC5466 is known to be disrupting \citep{Od04,Be06} and its
orbit takes it to within a few kpc of the solar neighbourhood
\citep[Fellhauer, private communication; see also][]{Fe07}. This,
along with the fact that the \sko a stars have similarly low
metallicity, indicates that they could well be tidal debris from
NGC5466.  However, the fact that there are four clusters located in
the same region of the ($\jz,\jperp$) plane raises the intriguing
possibility that all could be associated to a single major accretion
event. Previous authors have questioned whether subsets of these four
clusters may be associated \citep[e.g.][]{Di99,Pa02,Ma04}, but it
appears that no one to date has attempted to explicitly associate all
four clusters.

The uncertainties on the velocities of these clusters varies; NGC6205
and NGC7089 have reasonably well constrained angular momenta (with
error on $\jz$ and $\jperp$ of between 200 and 400 $\jun$), while for
NGC5466 and NGC6934 this is less well constrained (with error of
between 400 and 900 $\jun$). This is due to the uncertainties in the
proper motion combined with the large distances to these two clusters
($\sim$7 to $\sim$15 kpc). However, although the error for NGC5466
seems large, its proper motion is confirmed by \citet{Fe07} and so its
location in the ($\jz,\jperp$) plane is robust.

Note that a similar retrograde feature has been identified by
\citet[][see also Brook et al. 2003]{Di02} from the sample of
\citet{Ch00}. This feature, which they postulated could be debris torn
from the system that once contained $\omega$ Centauri, was observed as
an excess of stars with small retrograde velocities. However, as can
be seen by comparing our Fig. \ref{fig:skoa} to fig. 15 of
\citet{Ch00}, the stars in our \sko a feature are not compatible,
Although our stars indeed have small retrograde velocities, the
$\jperp$ velocities of \citet{Ch00} are significantly smaller than
ours and no more than one of their stars lies in our \sko a region.

\subsection{\sko b, \sko c}
\label{sec:skobc}

It is clear from Fig. \ref{fig:j_resid2} that there are a significant
number of more distant stars ($2.5 < \D < 5$ kpc) with large values of
$\jperp$. To establish whether these stars are clumped, we must
investigate their spatial distribution along the stripe. As our 
sampling is very inhomogeneous, we are actually interested in
identifying locations on the stripe where the fraction of high
$\jperp$ objects is anomalous, rather than the absolute number.

Since our field of view consists of a narrow 2.5$^\circ$ wide stripe
at constant declination, streams will typically cut through the stripe
and be localised in right ascension $\alpha$. Therefore, we analyse
our data by looking at $2^\circ$ sections in $\alpha$ along the
stripe, concentrating only on bins which contain more than 30
subdwarfs. We calculate the fraction of stars in a bin with $\jperp >
2000 \jun$ (considering all stars with $\D<5$ kpc), and also the
expected fraction as predicted from our smooth model. When we carry
out the calculation of the binomial probability, we take into account
the fact that there are 21 bins with more than 30 stars, i.e. we
determine the probability that any of these 21 bins has the observed
fraction of high $\jperp$ stars. The smooth model predicts a factor of
$\sim 1.9$ fewer high $\jperp$ stars in total compared to the observed
fraction (60/1717). This indicates the presence of non-Gaussian tails
in the velocity distributions and it is these outliers that could
harbour potential accretion remnants.

There are two locations on the stripe where the binomial probability
of obtaining the observed fraction of high $\jperp$ stars is less than
$\sim$0.05 and for which there are more than three such stars. We call
these two clumps \sko b and \sko c. Candidate members of these
overdensities are given in Table \ref{tab:sko}.

The clump \sko b is located in the range $46^\circ<\alpha<48^\circ$,
where we find 8 of 57 stars have $\jperp > 2000\jun$. Our smooth model
predicts that we should see only 2.1 stars and the corresponding
binomial probability is 0.03.  Various properties of the \sko b stars
are shown in Fig. \ref{fig:skob}, from which we see that there are a
significant number of stars in this $\alpha$ range at distances
between 3 and 5 kpc.  It is intriguing to note that most of these
objects have retrograde velocities ($\jz > 0$).  There is a clear
clump of three stars located in a small region of the
$(\vphi,\vth)$ plane\footnote{
Note that if these stars were debris from an accreted satellite, one might
naively expect them to also have the same values of $\vr$. However, as
can be seen from the modelling of satellite disruption for the H99
stream, this is not necessarily the case; although the model presented
in \citet{Ke07} is localised in $(\vz,\vphi)$-space, it has an
extended $\vR$ distribution.}
around (89, -225) $\kms$.  Furthermore, these three stars have similar
metallicities ($\langle\feh\rangle = -1.5$ dex) and distances (see
Table \ref{tab:sko}). Given that these stars are close to our limiting
distance of 5 kpc, it is worth looking at those stars which lie
outside this distance cut.  So in Fig. \ref{fig:skob}, we also 
include stars in this $\alpha$ range with distances up to 8 kpc.
Although their velocities are less certain, there are sufficient stars
to warrant their inclusion. Clearly, these more distant stars
reinforce the significance of the potential overdensity. The clump in
the $(\vphi, \vth)$ plane is even more pronounced, with six stars now
lying in this region. All six stars are located in a loose clump in
the $(\jz,\jperp)$ plane, with $\jz\in(670,1660)\jun$ and
$\jperp\in(2430, 3250)\jun$. Note that these additional three stars
also have similar metallicities to those with $D<5$ kpc. If we loosen
the cut on the SSPP flag as above (see Section \ref{sec:skoa}), then
we do not gain any additional potential members for this overdensity.

The second clump, \sko c, is located at $28^\circ<\alpha<30^\circ$ and
has 7 of 69 stars with $\jperp>2000\jun$. The smooth model predicts
only 1.9 stars (with corresponding binomial probability of 0.07). The
properties of these stars are illustrated in
Fig. \ref{fig:skoc}. Unlike \sko b, there is no obvious clumping in
velocity space. However, of note is the pair of two metal-poor stars
which lie at very similar distances ($\D \approx 4.1, 4.6$ kpc, $\feh
\approx -2.6, -2.6$ dex) and are relatively close to each other on the
sky (separated by half a degree). These two stars are located in close
proximity in the $(\vphi, \vth)$ plane, but their angular
momenta are not particularly close ($\jperp \approx 2000\pm460,
2890\pm730 \jun$). This means that the chance of these two stars
belonging to a kinematic overdensity is perhaps unlikely. Despite
this, their case is strengthened to some extent when we include stars
with less secure parameters; there is one additional metal-poor star
located in the $(\vphi, \vth)$ plane at (162, -202) $\kms$ with $\D
\approx 6.6$ kpc and $\feh \approx -2.5$ dex. As well as being more
distant than our standard cut of 5 kpc, there is also a cautionary
flag raised in the SSPP. The flag records that the star may be
exhibiting a strong CH $G$-band (around 4300 \AA) relative to what is
expected for a `normal' star \citep{Le08a} and implies that the
spectral parameters should be treated with caution. It lies very close
on the sky to the pair of metal-poor stars previously mentioned, but
given the various uncertainties it is far from clear whether we can
claim that it is associated to the pair of metal-poor stars in this
region. Therefore, in conclusion, the validity of this feature is
rather tentative and it could simply illustrate that we have reached
the limits of this method.

\section{Conclusions}

This paper has presented a catalogue of $\sim1700$ halo subdwarfs
within a heliocentric distance of $\sim5$ kpc.  Our analysis is
restricted to a particular patch of sky, Stripe 82 of the Sloan
Digital Sky Survey (SDSS), and exploits the light-motion curve
catalogue constructed by \citet{Br08}.  The exceptional
precision of the SDSS photometry allows us to determine distances with
an uncertainty of $\sim10$ per cent from a photometric parallax
relation. As a consequence, we obtain high-quality tangential
velocities (with typical errors of $30-50\kms$) using the proper
motions derived from the light-motion curves. Radial velocities are
obtained from SDSS spectra. So, this sample of halo stars has full
six-dimensional phase-space information, allowing us to probe in
detail the kinematical structure of the stellar halo.  As our data are
restricted to Stripe 82, we may be susceptible to spatial variations
present in the halo population. However, even if the halo population
is not well-mixed, our $\sim $250 square degree field-of-view is
sufficiently large to overcome any small-scale effects.

The size of our sample allows us to determine the orientation and
semi-axes of the velocity ellipsoid of the stellar halo to excellent
accuracy. The velocity ellipsoid is almost precisely aligned in the
spherical polar coordinate system, indicating that the total potential
must be nearly spherical~\citep{Sm09}. The velocity dispersions are
$(\sigma_r,\sigma_\phi,\sigma_\theta)$ = (143 $\pm$ 2, 82 $\pm$ 2, 77
$\pm$ 2) $\kms$.  The values of the dispersions are significantly
smaller than previous estimates, although the ratios of the
dispersions are in good agreement. A simple distribution function that
matches the kinematic data is constructed -- and it suggests that the
density of the stellar halo is only mildly flattened and falls off
with distance like $\rho \sim r^{-3.8}$.

The stellar halo exhibits no net rotation.  The velocity dispersions
are reasonably well fit by a Gaussians, with the exception of the
$\vphi$ component which is asymmetric. We believe that this is a real
effect. It appears to be more pronounced for metal rich stars in our
sample. None of the kinematic substructures are able to account for
this asymmetry.

On the issue of substructure, we confirm the presence of the stream
identified by \citet{He99}. Using the angular momentum
components ($\jz, \jperp$), we find a number of other potential
substructures, which we label Sloan Kinematic Overdensities (\sko
s). Two of these new features are particularly striking, containing a
number of stars that are localised in both kinematics and
metallicity. The metal-poor overdensity \sko a ($\langle\feh\rangle
\approx -2.1$ dex) appears to be coincident in the ($\jz,\jperp$)
plane with an association of four globular clusters (NGC5466, NGC6934,
NGC7089/M2 and NGC6205/M13), suggesting that they may have been part
of the same accretion event. If so, then this implies that the
progenitor must have been a large satellite, similar in size to
Fornax. We are currently investigating the orbits of these stars and
clusters and plan to test this hypothesis using simulations.

\section*{Acknowledgments}

The authors wish to thank Mario Juri\'c, Matthew G. Walker and an
anonymous referee for advice and guidance, as well as Jeff Munn and
{\L}ukasz Wyrzykowski for help with the Light-Motion Curve
Catalogue. MCS acknowledges support from the STFC-funded ``Galaxy
Formation and Evolution'' program at the Institute of Astronomy,
University of Cambridge.

Funding for the SDSS and SDSS-II has been provided by the Alfred
P. Sloan Foundation, the Participating Institutions, the National
Science Foundation, the U.S. Department of Energy, the National
Aeronautics and Space Administration, the Japanese Monbukagakusho, the
Max Planck Society, and the Higher Education Funding Council for
England. The SDSS Web Site is http://www.sdss.org/.

The SDSS is managed by the Astrophysical Research Consortium for the
Participating Institutions. The Participating Institutions are the
American Museum of Natural History, Astrophysical Institute Potsdam,
University of Basel, Cambridge University, Case Western Reserve
University, University of Chicago, Drexel University, Fermilab, the
Institute for Advanced Study, the Japan Participation Group, Johns
Hopkins University, the Joint Institute for Nuclear Astrophysics, the
Kavli Institute for Particle Astrophysics and Cosmology, the Korean
Scientist Group, the Chinese Academy of Sciences (LAMOST), Los Alamos
National Laboratory, the Max-Planck-Institute for Astronomy (MPIA),
the Max-Planck-Institute for Astrophysics (MPA), New Mexico State
University, Ohio State University, University of Pittsburgh,
University of Portsmouth, Princeton University, the United States
Naval Observatory, and the University of Washington.

\appendix

\section{Potential Contaminants}
\label{app:contam}

There are three main sources of contamination: white dwarfs, disc
dwarfs and background giants. We discuss each in turn and conclude
that our contamination level is less than $\sim 1$ per cent.

\subsection{White dwarfs}

In a RPM diagram, the white dwarfs are fainter and bluer than the halo
subdwarfs. Their presence in Fig. \ref{fig:rpm} is only just
discernible as the cloud around ($(g\!-\!i), H_r$) $\approx$ (-0.4,
17). Although the density of white dwarfs in this figure is clearly
small, it is possible that white dwarfs with small proper motions
enter our subdwarf region. As stated in Section \ref{sec:spectra}, we
require that the SSPP flag is set to `nnnn'. This flag includes a
check for candidate white dwarfs \citep{Le08a} and so our sample
should be free from such contamination.

As a check, we cross-matched our subdwarf sample with the catalogue of
spectroscopically confirmed white dwarfs from the fourth data release
(DR4) of the SDSS~\citep{Ei06}. This catalogue was constructed via
both automated criteria and visual inspection and so should be
reasonably complete. We find that none of their white dwarfs overlap
with our subdwarf sample. Although the \citet{Ei06} catalogue contains
only DR4 data, we know that $\sim 60$ per cent of our subdwarf spectra
were included in DR4 and so, assuming the fraction of white dwarfs
does not vary with SDSS data releases, we can expect $\la 1$ white
dwarf to contaminate our subdwarf sample.

\subsection{Disc dwarfs}

Although our RPM cut in Section \ref{sec:rpm} rejects stars with
kinematics similar to the Sun (hence retaining mainly fast moving halo
stars), it is possible that some disc stars could enter our
sample. Contamination by thin-disc stars is negligible due to two
factors. First, the small scale-height ($\sim 200$ pc) of the thin
disc means that our sample, which goes out to a few kpc below the
plane, should have minimal contamination. Second, the thin disc is
cold so that it is highly unlikely that a thin disc star would have
large enough proper motion to pass our RPM cut. However, it is
possible that there may be some contamination from the hotter
thick-disc component. Previous studies based on similar RPM cuts have
found levels of disc contamination to be of order two per cent
\citep{Go07}.

To quantify the contamination from disc dwarfs in our sample we
analyse the kinematics of the stars. We calculate the probability that a star
belong to a particular population (i.e. thin disc, thick disc or halo)
by comparing its velocity to a toy model constructed using Gaussian
distributions for each component of the velocity, taking into
account the different scale-heights and local density
normalisations. For the scale-heights and disc kinematics we follow
\citet{Sm07}, with the exception of the mean rotation velocity of the
thick disc which we model using the parameters determined from
\citet{Gi06}, i.e. $\vphi = -195 + 30\,|${\it z}$|\,\kms$. We model
the halo component using the kinematics of \citet{Ke07} and assume
that the halo density does not vary significantly within the volume
probed. As expected, we find that none of our stars are compatible
with the thin disc. For the thick disc we analyse the stars with
$\feh>-1.5$ dex, since this is the three-sigma lower limit on the
thick-disc metallicity according to \citet{So03}. Although there are
no stars in this $\feh$ range with thick-disc membership probabilities
of greater than 0.36, there are six stars with probabilities of
between 0.2 and 0.36. If we add up the thick-disc probabilities of all
stars in this $\feh$ range we can estimate the total level of
contamination. Note that this will probably be an overestimate because
the kinematics of the thick disc overlap those of the halo, which
implies that even a pure halo sample will be estimated to have a
non-zero level of contamination. Using this technique we estimate that
the thick-disc contamination should be no more than 10.2 stars,
i.e. less than one per cent. This should have little bearing on our
results. As a test we repeated the analysis of Section
\ref{sec:kinematics} using only stars at distances of more than 2.5
kpc from the Galactic plane and found that global kinematic properties
were unchanged.


\subsection{Background Giants}

Another possible source of contamination is from faint background
giants. However, given the RPM cut, we see that a star must have a
sufficiently high $\mu$ to shift it into the subdwarf regime. For
example, a stationary distant star with $r=19$ mag and a spurious $\mu
= 4$ \masyr would have $H_r = 12$ mag. If the star was blue enough
then it could conceivably pass the $H_r$ cut. However, in our final
subdwarf sample we find that the star with the least significant
proper motion is inconsistent with $\mu=0$ at the $2.5\sigma$ level,
which means contamination from distant (and hence stationary) stars is
negligible.

\section{Turn-off correction for photometric parallax relation}
\label{app:turn-off}

\begin{figure}
\plotone{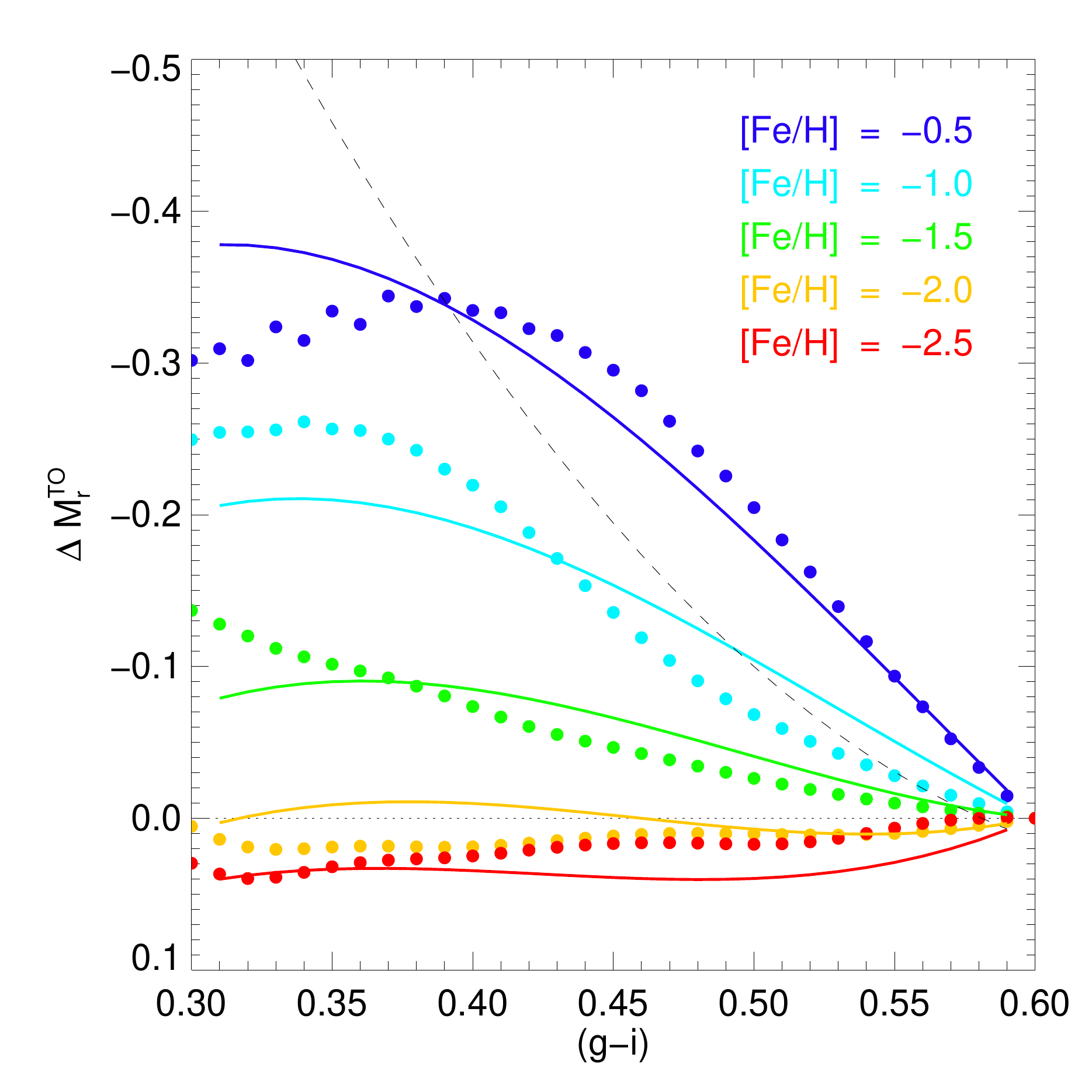}
\caption{Our correction to the photometric parallax relation due to
the main-sequence turn-off. The points denote mean magnitudes from the
stellar models and the corresponding solid curves denote the
polynomial fit given in equation \ref{eq:turn-off_app}, with
\feh\ increasing from -2.5 (bottom line) to -0.5 (top line). The
dashed line corresponds to the turn-off correction of \citet{Iv08}.}
\label{fig:turn-off}
\end{figure}

We estimate distances using a photometric parallax relation based on
\citet{Iv08}, who construct a relation using data from a number of
globular clusters. They first determine a $(g\!-\!i) - M_r$
colour-magnitude sequence for stars on the main-sequence by
identifying the colour of the main-sequence turn-off for these
clusters (i.e. the point at which the slope of the colour-magnitude
relation becomes vertical) and discarding all data within 0.05 mag of
the turn-off. Once they have this relation, they then devise a
correction to account for the presence of the main-sequence turn-off,
which they base on the sequence for the cluster M13.

However, the morphology of the colour-magnitude relation around the
turn-off region will clearly be dependent on both metallicity and age,
and so we would like to construct a correction which incorporates
these effects. In order to do this, we use the stellar models of
\citet{Do08}, taking isochrones with ages from 1 to 15 Gyr (in steps
of 0.5 Gyr) and \feh\ from -0.5 to -2.5 (in steps of 0.5
dex). Following the compilation of \citet{Ve04}m we choose $[{\rm
\alpha/Fe}]=0.3$, which is representative for our halo \feh\ range.

We take each set of isochrones and, in the same manner as
\citet{Iv08}, we shift each sequence so that it has $M_r=0$ for
$(g\!-\!i)=0.6$. Then, for a given {\feh}, we calculate mean $M_r$ as
a function of $(g\!-\!i)$ in the range $0.3<(g\!-\!i)<0.6$,
considering only model data up to the main-sequence turn-off. Since we
know the approximate age distribution of the halo population, we
calculate a weighted mean using a Gaussian prior with mean and sigma
of 10 and 2 Gyr, respectively. \footnote{One could also incorporate a
  prior based on the initial mass function or, equivalently, the
  luminosity function. However, we investigated this and found it had
  a negligible effect on our results.}  We then calculate the offset
between the mean model magnitude and the uncorrected relation of
\citet{Iv08}, i.e. that given in equations (A1-A5). This is shown in
Fig. \ref{fig:turn-off}, where the effect of the turn-off correction
is to increase the brightness of stars compared to the uncorrected
relation.

This data is then fit with a third-order polynomial, resulting in the
following equation which is applicable for stars in the range
$0.3<(g\!-\!i)<0.6$ and $-2.5<\feh<-0.5$,
\beq
\label{eq:turn-off_app}
\Delta M_r^{TO} = 
a_0\,x + 
a_1\,x\,y +
a_2\,x^3 +
a_3\,x^2\,y +
a_4\,x\,y^2,
\eeq
where $x = (g\!-\!i) - 0.6$, $y = \feh$, and $a_0 = 2.87,\: a_1 =
2.25,\: a_2 = -9.79,\: a_3 = 2.07,\: a_4 = 0.31$. Note that we do not
include terms depending solely on $y$ since we require the relation to
go through $\Delta M_r^{TO}=0$ for $(g\!-\!i)=0.6$. We also discard
the $x^2$ term since this has no effect on the fit. Although the fit
is far from perfect, given the overall uncertainties in the method we
believe this should provide a reasonable approximation.

In order to avoid any divergent behaviour in this relation we do not
extrapolate equation (\ref{eq:turn-off_app}) beyond
$\feh=(-2.5,-0.5)$. For stars with metallicities outside this range, we
use the relation at the limit of our allowed range (i.e. $-2.5$ dex
for the metal-poor stars and $-0.5$ dex for the metal-rich stars).
This has very little bearing on our results: although we have around
30 stars with $0.3 < (g\!-\!i) < 0.6$ and $\feh<-2.5$, the relation
converges at the metal-poor end and so the $\feh=-2.5$ relation should
provide a good approximation; from Fig. \ref{fig:turn-off} it can be
seen that the relation does not converge at the metal-rich end, but
this is of little consequence since there are fewer than 10 stars with
$0.3 < (g\!-\!i) < 0.6$ and $\feh\ > -0.5$.

Clearly the uncertainties on the parallax relation will be larger in
this colour range due to the scatter in this turn-off correction. To
estimate the uncertainty, we calculate the standard deviation in $M_r$
when we calculate the mean. The scatter varies as a function of colour
and metallicity, but if we take the relation with the largest scatter
(corresponding to $\feh=-0.5$ dex) we find the following relation,
\beq
\delta\left(\Delta M_r^{TO}\right) = 0.39 - 0.65\,(g\!-\!i).
\eeq
When estimating distances for our stars we add this uncertainty in
quadrature to the other sources of error.

\end{document}